\documentclass[a4paper,11pt]{article}
\usepackage{hyperref}
\usepackage{geometry}
\usepackage{amsmath,amssymb}
\usepackage{pdflscape}
\usepackage{cancel}
\usepackage{subfigure}
\usepackage{mathrsfs}
\usepackage{graphicx}
\usepackage{hepunits}
\usepackage{wasysym}
\usepackage{feynmf}
\usepackage{multirow}
\usepackage[utf8x]{inputenc}
\usepackage{paralist}

\usepackage{changepage}
\usepackage[numbers,sort&compress
,comma]{natbib}

\def\lsim{\;\raise0.3ex\hbox{$<$\kern-0.75em\raise-1.1ex\hbox{$\sim$}}\;}
\def\gsim{\;\raise0.3ex\hbox{$>$\kern-0.75em\raise-1.1ex\hbox{$\sim$}}\;}

\newcommand\tev{{\rm TeV}}
\newcommand\CO{{\cal O}}
\newcommand\simge{\mathrel{%
   \rlap{\raise 0.511ex \hbox{$>$}}{\lower 0.511ex \hbox{$\sim$}}}}

\author{{\large Johann Brehmer$^{1}$,
Gustaaf Brooijmans$^{2}$,
Giacomo Cacciapaglia$^{3}$,
Adrian Carmona$^{4}$,} \\
{\large R.~Sekhar Chivukula$^{5}$,
Antonio Delgado$^{6}$,
Florian Goertz$^{4}$,
JoAnne L.~Hewett$^{7}$,} \\
{\large Andrey Katz$^{4,8}$,
Joachim Kopp$^{9}$,
Kenneth Lane$^{10}$,
Adam Martin$^{6}$,} \\
{\large Kirtimaan Mohan$^{5}$,
David M.~Morse$^{11}$,
Marco Nardecchia $^{12}$,
Jose Miguel No$^{13}$,} \\
{\large Alexandra Oliveira$^{14}$,
Chris Pollard$^{15}$,
Mariano Quiros$^{16}$,
Thomas G. Rizzo$^{7}$,} \\
{\large Jose Santiago$^{17}$,
Veronica Sanz$^{13}$,
Elizabeth H. Simmons$^{5}$,
Jamie Tattersall$^{18}$
}\\ [0.4cm]
{\footnotesize $^{1}$ Institut f\"ur Theoretische Physik, Universit\"at Heidelberg, Germany} \\
{\footnotesize $^{2}$ Nevis Laboratories, Columbia University, New York, NY 10027, USA} \\
{\footnotesize $^{3}$ Universit\'e de Lyon, F-69622 Lyon, France; Universit\'e Lyon 1, Villeurbanne} \\
{\footnotesize and CNRS/IN2P3, UMR5822, Institut de Physique Nucl\'eaire de Lyon, F-69622 Villeurbanne Cedex, France}\\
{\footnotesize $^{4}$ Theory Division, CERN, 1211 Geneva 23, Switzerland} \\
{\footnotesize $^{5}$ Department of Physics and Astronomy, Michigan State University, East Lansing, MI 48824, USA} \\
{\footnotesize $^{6}$ Department of Physics, University of Notre Dame, Notre Dame, IN 46556, USA} \\
{\footnotesize $^{7}$ SLAC National Accelerator Laboratory, Menlo Park, CA 94025, USA} \\
{\footnotesize $^{8}$ Universit\'e de Gen\`eve, Department of Theoretical Physics and Center for Astroparticle Physics (CAP),}\\
{\footnotesize  CH-1211 Geneva 4, Switzerland} \\
{\footnotesize $^{9}$ PRISMA Cluster of Excellence, 55099 Mainz, Germany, and } \\
{\footnotesize Mainz Institute for Theoretical Physics, Johannes Gutenberg-Universit\"{a}t Mainz, 55099 Mainz, Germany}  \\
{\footnotesize $^{10}$ Department of Physics, Boston University, Boston, MA 02215, USA} \\
{\footnotesize $^{11}$ Department of Physics, Northeastern University, Boston MA, USA} \\
{\footnotesize $^{12}$ DAMTP, University of Cambridge, Wilberforce Road, Cambridge CB3 0WA, United Kingdom} \\
{\footnotesize $^{13}$ Department of Physics and Astronomy, University of Sussex, BN1 9QH Brighton, UK} \\
{\footnotesize $^{14}$ Dipartimento di Fisica e Astronomia and INFN, Sezione di Padova, I-35131 Padova, Italy} \\
{\footnotesize $^{15}$ School of Physics and Astronomy, University of Glasgow, G12 8QQ Glasgow, UK} \\
{\footnotesize $^{16}$ Institut de Fisica d'Altes Energies (IFAE), The Barcelona Institute of  Science and Technology (BIST), } \\
{\footnotesize Instituci\'o Catalana de Recerca i Estudis Avan\c{c}ats (ICREA), Campus UAB, 08193 Barcelona, Spain,} \\
{\footnotesize and ICTP-SAIFR \& Instituto de F\'isica Te\'orica, Universidade Estadual Paulista, S\~ao Paulo, Brazil} \\
{\footnotesize $^{17}$ CAFPE and Dpto Fisica Teorica y del Cosmos, University of Granada, E-18071, Granada, Spain} \\
{\footnotesize $^{18}$ Institut f\"ur Theoretische Teilchenphysik und Kosmologie, RWTH Aachen, Germany}}

\title{The Diboson Excess:\\
Experimental Situation and Classification of Explanations\\
A Les Houches Pre-Proceeding}

\date{\today}
\begin{document}

\maketitle
\newpage
\begin{flushright}
CERN-PH-TH-2015-297\\[0.2cm]
\end{flushright}
\begin{abstract}
We examine the `diboson' excess at $\sim 2$\, TeV seen by the LHC experiments in various channels. We provide a comparison of the excess significances as a function of the mass of the tentative resonance and give the signal cross sections needed to explain the excesses.
We also present a survey of available theoretical explanations of the resonance, classified in three main approaches. Beyond that, we discuss methods to verify the anomaly, determining the major properties of the various surpluses and exploring how different models can be discriminated.
Finally, we give a tabular summary of the numerous explanations, presenting their main phenomenological features.
\end{abstract}

\tableofcontents
\newpage

\section{Introduction}
The LHC has pushed particle physics to a new energy frontier. After the first successful run at 8 TeV when the Higgs was discovered, the machine restarted this year with collisions at 13 TeV. This new run will serve to study the Higgs properties to a much better precision that we already know but it will also confirm or disregard some possible new physics signals coming from excesses in different channels with respect to the expected number of events.

Several of those excesses come from the search for heavy resonances around 2 TeV, and the fact that they are in several channels and appearing in both ATLAS and CMS analyses has driven lots of attention in the community. The possible signal could be interpreted as a heavy bosonic resonance decaying into two SM electroweak vector bosons. There have been many theoretical papers trying to explain the signal within different models and there have also been some attempts to combine the different channels into one single significance plot.

The aim of this paper is to summarize in a single document both the experimental and the theoretical situation of the 'diboson' excess in preparation for the data coming from the LHC 13 TeV run. Should the excess be confirmed the reader could easily use this document as a first point to check which one of the possible approaches are more likely to be able to explain not only the existing excesses but any other  that may come in different channels. By having a comprehensive analysis of the different ideas we hope to present a clear explanation of the situation.

The paper is organized as follows. The different experimental signals are summarized in section 2. Section 3 contains a summary of the different possible resonances that could explain the excesses. We will discuss the phenomenology for run 2 in section 4, section 5 is devoted to an overview of the different models and finally our conclusions are presented in section 6.

\section{Experimental Results}
\label{sec:Exp}


The primary reason the ``diboson'' excess is of interest is that there are excesses in multiple 
analyses searching for heavy resonances in the region between 1.5 and 2.5 TeV.  These excesses 
are not limited to the diboson decay channels, so any multi-channel combination would require 
significant assumptions about the underlying model, leading to an inherent bias in the ensuing 
conclusions.  
Instead, the results are presented 
in two ways: a) a comparison of the excesses as a function of mass in units of standard 
deviations, and b) a comparison of the signal cross sections needed to lead to these excesses, also
as a function of resonance mass.
As this was finalized, an informal ATLAS--CMS combination of individual channels was released~\cite{Dias:2015mhm}.

\subsection{Results Considered}

The following experimental results are considered:
\begin{compactitem}
\item Diboson resonance searches that do not include Higgs bosons: 
 ATLAS and CMS  searches in the all-hadronic~\cite{Aad:2015owa,Khachatryan:2014hpa}, 
 $\ell\nu$ (where $\ell$ denotes an electron or muon) plus hadrons~\cite{Aad:2015ufa,Khachatryan:2014gha} 
 and $\ell\ell$ plus hadrons~\cite{Aad:2014xka,Khachatryan:2014gha} final states;
\item ATLAS and CMS $WH$ resonances searches in the $\ell\nu b \bar{b}$ final state~\cite{Aad:2015yza,CMS-PAS-EXO-14-010};
\item An ATLAS $ZH$ resonance search in the $\ell\ell b \bar{b}$ and $\nu\nu b \bar{b}$ final states~\cite{Aad:2015yza};
\item A CMS $WH$ and $ZH$ resonance search in the all-hadronic final state~\cite{Khachatryan:2015bma};
\item A CMS $ZH$ resonance search in the $\tau\tau$ plus hadrons final state~\cite{Khachatryan:2015ywa};
\item A CMS $W_R$ boson search in the $\ell\ell$ plus hadrons~\cite{Khachatryan:2014dka};
\item ATLAS and CMS searches for dijet resonances~\cite{Aad:2014aqa,Khachatryan:2015sja};
\item ATLAS and CMS dilepton resonance searches~\cite{Aad:2014cka,Chatrchyan:2012oaa}.
\end{compactitem}

\subsection{Comparison of Excess Significances}

\begin{figure}[htbp]
\centering
\subfigure[]
{
\includegraphics[width=0.55\textwidth]{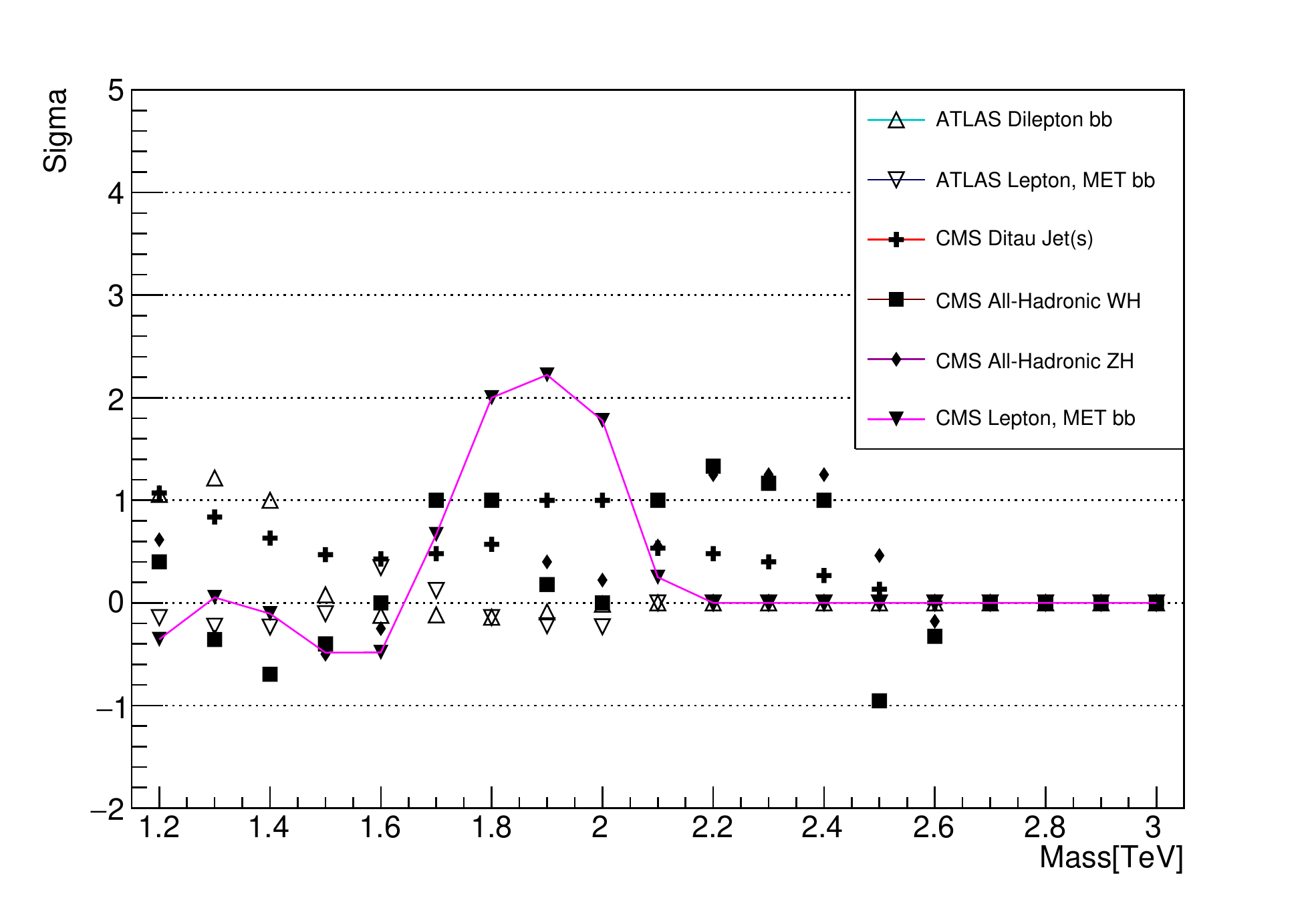}
}
\subfigure[]
{
\includegraphics[width=0.55\textwidth]{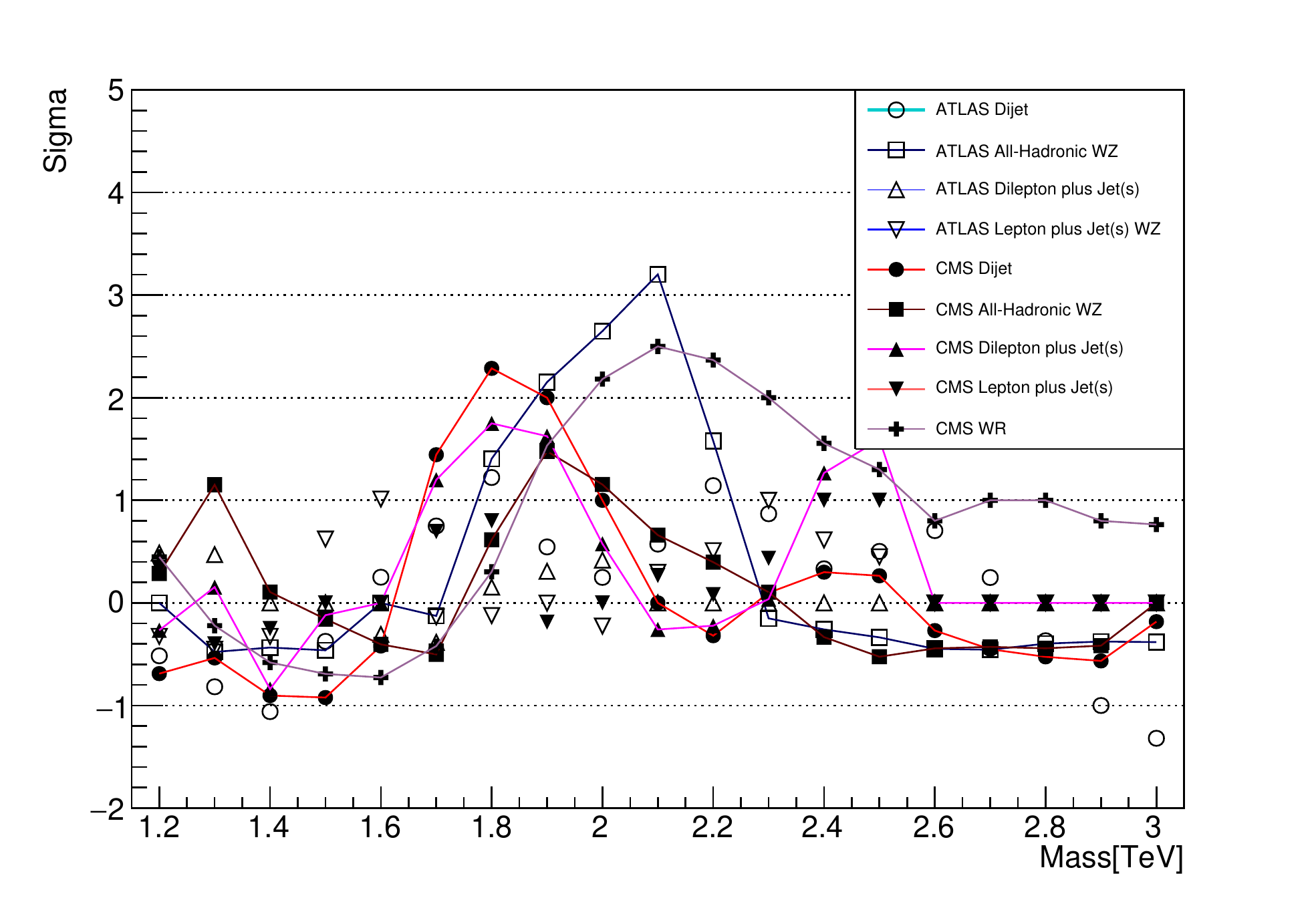}
}
\subfigure[]
{
\includegraphics[width=0.55\textwidth]{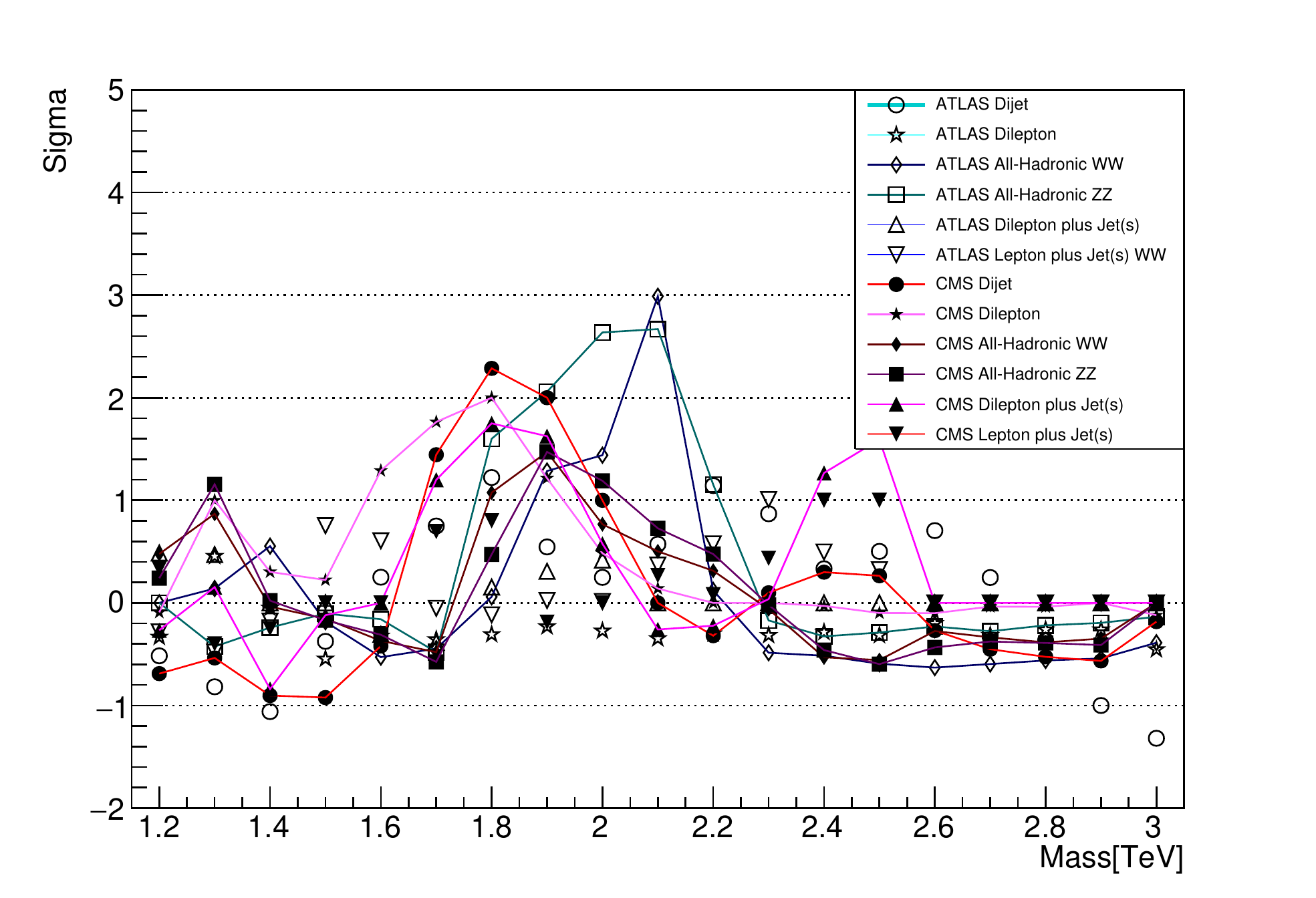}
}
\caption{Comparison of the excess magnitudes in high mass resonance searches in units of standard deviations for
(a) decays including a Higgs boson, and (B,c) decays without Higgs bosons compatible with a charged (b) and a 
neutral (c) resonance.  Channels which exceed a 1.5 $\sigma$ excess are shown with lines, others without.}
\label{fig:sigmas}
\end{figure}

To extract this information, the limit plots from the different results are digitized: the observed, 
expected, and expected plus one and two $\sigma$ limits are recorded for resonance mass values between 
1.2 and 3.0 TeV in 100 GeV steps.  To calculate the excesses in units of $\sigma$, simple approximations 
are used: if the excess lies between one and two $\sigma$, a linear interpolation between the expected plus one and 
two $\sigma$ values is performed; if the excess is less than one (or larger than two) $\sigma$, a linear 
interpolation (extrapolation) using the expected and expected plus two $\sigma$ values is performed.
This method attempts to take into account that the slope of a gaussian is steeper between one and 
two $\sigma$ than elsewhere.  While the expected limit distribution for a given resonance mass is not necessarily
gaussian, for this comparison the approximation is of sufficient quality.
The results are shown in Fig.~\ref{fig:sigmas} separately for decays including a Higgs boson, and
decays without Higgs bosons, compatible with either a charged or neutral resonance.
For decays with a Higgs boson, only the CMS $\ell\nu b \bar{b}$ search shows a significant excess, between 
1.7 and 2.1 TeV.  For charged and neutral resonances, overlapping significant excesses are found in the 
region between 1.8 and 2.0 TeV.

\subsection{Signal Cross Sections}

The signal cross sections needed to produce the observed excesses are calculated based on the observed, 
expected, and expected plus one and two $\sigma$ limits.  
For the dijet resonances, the limit 
plots include an acceptance factor, and the results are divided by the acceptance factor for $W'$ bosons
given by ATLAS.  (Acceptance values for $W'$ bosons from CMS were not found, but for other signals the values
for ATLAS and CMS agree well.)  For the CMS dilepton resonance search, the limit is given as a ratio
to the $Z$ boson cross section times branching ratio to electrons and muons.  To convert to fb, this 
is assumed to be equal to 2 nb.

\begin{figure}[htbp]
\centering
\includegraphics[width=0.65\textwidth]{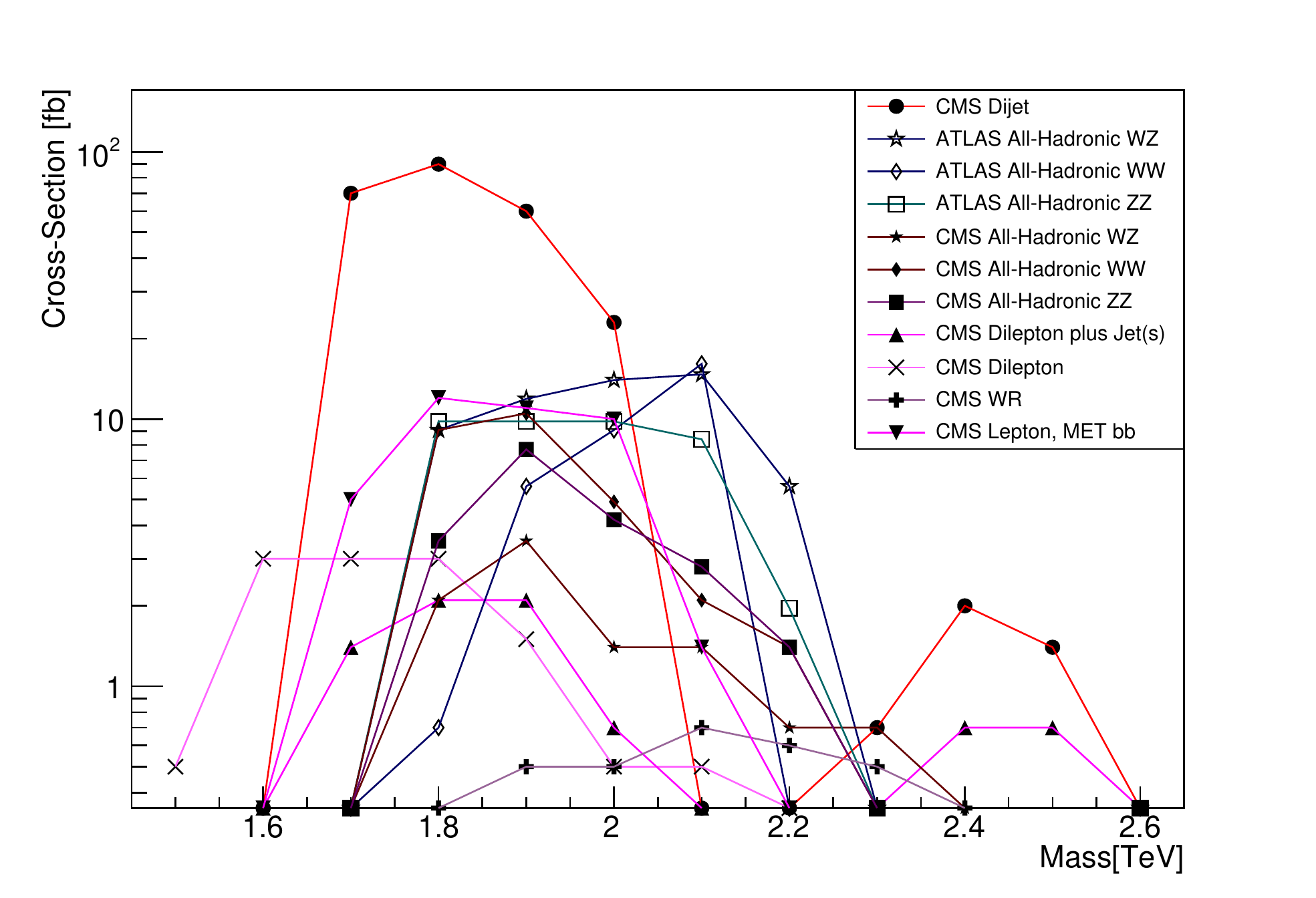}
\caption{Estimated signal cross sections for searches with excesses larger than 1.5 $\sigma$.}
\label{fig:xs}
\end{figure}

To evaluate the signal cross sections, each mass point is initially treated as a single
counting experiment.  Signal and background are each assumed to be affected by a single uncertainty, and 
these uncertainties, as well as the signal and background cross sections are adjusted to reproduce the 
values obtained from the limit plots.  The limits are calculated using the CL$_s$ method with the profile 
likelihood test statistic (as done by the experiments). The background uncertainty 
drives the differences between the 
expected and expected plus one and two $\sigma$ limits.  In this simplified approach, it is often 
impossible to make it small enough to match the values in the limit plots.  Tests indicate this has a 
negligible impact on the signal cross section needed to reproduce the excess.  To account for
the use of the full invariant mass distribution by the experiments, the obtained signal cross sections
are multiplied by a factor 0.7.  This is obtained from an ATLAS study comparing the sensitivity to 
dilepton resonances using the full invariant mass distribution with that obtained from counting 
experiments in narrow mass windows~\cite{Aad:2009wy}.  The resulting signal cross sections agree 
reasonably well with those obtained when using the number of observed events, signal efficiencies and 
acceptances in the few cases where this information is available.
An uncertainty of 30\% should cover the limitations of the method used.
The results are shown in Figure~\ref{fig:xs}.

\section{Theoretical Interpretation}

In this section, we summarize the major approaches for explaining the diboson excess in terms of extensions of the Standard Model.  Most of these involve production and decay of a single, relatively narrow $s$-channel resonance.  We begin by using a simplified model of such a resonance to convert an estimated signal cross-section into model-independent {\it lower} bounds on the branching ratios for the resonance that correspond to the production and decay modes.  For estimated cross-sections of 1 - 100 fb, we find that the vector-boson fusion production mode must be subdominant.  Next, we turn to a discussion of the three main classes of physics beyond the standard model that have been invoked to explain the diboson excess, namely, strongly-coupled scenarios such as composite Higgs models, modifications of the Standard Model (SM) gauge group, and extensions of the scalar sector. More exotic solutions beyond this classification will be summarized~in~Section~\ref{sec:overv}.

\subsection{Simplified Model Description of an $s$-channel Resonance}


	      There are many Beyond the Standard Model possibilities that may explain the diboson excesses discussed in this work. In most cases, the excess arises from the production and decay of a single, relatively narrow, $s$-channel resonance. In this context, a simplified model of the resonance allows us to convert any estimated signal cross section into model-independent constraints on the properties of the resonance. In particular, if resonance production occurs dominantly through a single process, we can obtain model-independent {\it lower} bounds on the product of the branching ratios corresponding to production and decay. Any resonance proposed to explain an excess in pairs of final-state weak bosons can certainly be produced via vector-boson fusion and can potentially be produced through $q\bar{q}$ annhilation and/or gluon fusion as well. The simplified model analysis shows immediately that, in order to explain the size of the excesses discussed here, vector boson fusion cannot be the dominant production mechanism.
	      
	The tree-level partonic production cross-section for an arbitrary 
	$s$-channel resonance at the LHC can be written \cite{Harris:2011bh,Agashe:2014kda}
	\begin{equation}
	\hat{\sigma}_{ij\to R\to xy}(\hat{s}) = 16 \pi \cdot {\cal N} \cdot
	\frac{\Gamma(R\to i+j) \cdot \Gamma(R\to x+y)}
	{(\hat{s}-m^2_R)^2 + m^2_R \Gamma^2_R} ~,
	\label{eq:resonance-sigma}
	\end{equation}
	where ${\cal N}$ is a ratio of spin and color counting factors
	\begin{equation}
	{\cal N} = \frac{N_{S_R}}{N_{S_i} N_{S_j}} \cdot
	\frac{C_R}{C_i C_j},
	\end{equation}
	where $N_S$ and $C$ count the number of spin- and color-states for initial state partons $i$ and $j$ and for the resonance $R$.	In the narrow-width approximation one can simplify this further using.
	\begin{equation}
	\frac{1}
	{(\hat{s}-m^2_R)^2 + m^2_R \Gamma^2_R}
	\approx \frac{\pi}{m_R \Gamma_R} \delta(\hat{s} - m^2_R)~.
	\end{equation}

Assuming that one production mechanism ($x+y\to R$), dominates, 
we can write down the signal cross-section for pp-collisions as follows.
	\begin{equation}
	\sigma_R(pp \to x+y) = \int_{s_{min}}^{s_{max}}d\hat{s}\,
	\hat{\sigma}(\hat{s}) \cdot \left[ \frac{d L^{ij}}{d\hat{s}}\right]~,
	\label{eq:simplest}
	\end{equation}
and hence
\begin{equation}
\sigma_R(pp \to x+y) = 16 \pi^2 {\cal N} \frac{\Gamma_R}{m_R} BR(R\to i+j) \cdot BR (R \to x + y)
\left[ \frac{d L^{ij}}{d\hat{s}}\right]_{\hat{s}=m^2_R}~.
\label{eq:real-simple}
\end{equation}
	Here $ \frac{d L^{ij}}{d\hat{s}}$ corresponds to the luminosity function for the $ij$ combination of partons\footnote{
	We calculate these parton luminosities using the {\tt CTEQ6L1}~\cite{Pumplin:2002vw} parton density functions, setting the factorization scale $q^2= \sqrt{\hat{s}}$.}.
	
	Using this expression, we can immediately derive model-independent bounds on the branching ratios
	of the resonance \cite{Chivukula:2016}.
	
	To illustrate this point, consider the possibility of a charged spin-one color-neutral vector
	resonance -- a technirho or a $W^\prime$ -- decaying to $W^\pm Z$. Such an object can be produced either via vector boson fusion or via $q\bar{q}$ (in this case primarily $u\bar{d}$ or $d \bar{u}$) annihilation. For vector boson fusion production, we see that the signal cross section (in this simplified model) is determined entirely by $BR(R \to W Z)$, which is bounded from above by 1; for $q\bar{q}$ production, on the other hand, the signal cross section is determined entirely by $BR(R\to q\bar{q}) \cdot BR(R\to WZ)$, which is bounded from above by $1/4$.
	
\begin{figure}[htbp]
\centering
\includegraphics[width=1.0\textwidth]{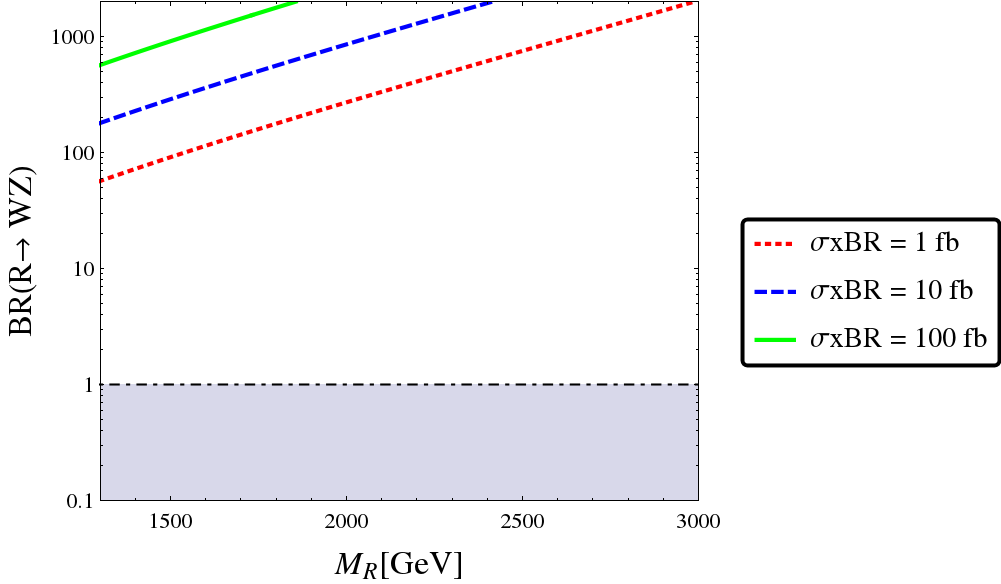}
\caption{Lower bounds on branching ratio $BR(R\to WZ)$ assuming production of an s-channel resonance $R$ with $\Gamma_R/m_R = 0.1$ via vector boson fusion, shown for three different values of $\sigma\times BR$. Since the lower bounds exceed the value 1, the VBF process cannot be the dominant production mode for a resonance R that is causing the observed diboson excess.}
\label{fig:simplified-vbf}
\end{figure}

\begin{figure}[htbp]
\centering
\includegraphics[width=1.0\textwidth]{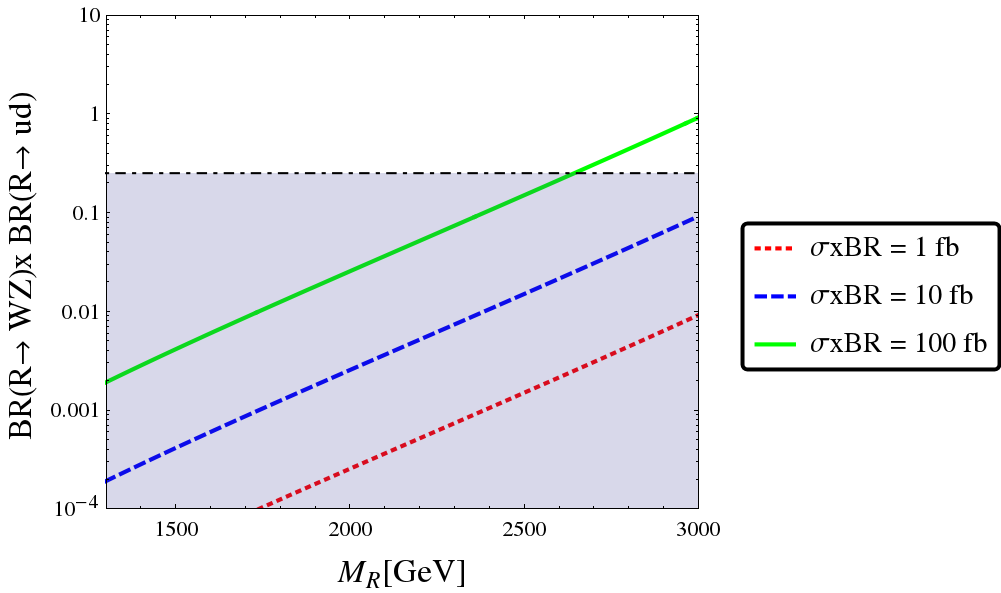}
\caption{Lower bounds on the product of branching ratios $\left[BR(R\to WZ)\times BR(R\to ud)\right]$ for production of an s-channel resonance $R$ with $\Gamma_R/m_R = 0.1$ via quark anti-quark annihilation, for three different values of $\sigma\times BR$.  The portions of the curves lying within the shaded region correspond to values of the product of branching ratios for which a $q\bar{q}$-produced resonance could be consistent with the observed diboson excess.}
\label{fig:simplified-qqb}
\end{figure}

	    In Figures~\ref{fig:simplified-vbf} and \ref{fig:simplified-qqb} we plot the {\it lower bounds} for the appropriate branching ratios corresponding to vector boson fusion (calculated in the effective $W$ approximation \cite{Dawson:1984gx}) and $q\bar{q}$ annihilation production assuming $\Gamma_R/m_R = 0.1$ and for signal cross sections of 1, 10, and 100 fb -- of order the excesses described in this work. Looking at Figure~\ref{fig:simplified-vbf}, we see that the lower bounds on $BR(R \to WZ)$ exceed 1; therefore, vector boson fusion must be only a subdominant production mode for any resonance responsible for the observed diboson excess. In contrast, from Figure~\ref{fig:simplified-qqb} we see that $q\bar{q}$ annhilation can be consistent with the observed excesses so long as the product of the branching ratios to $WZ$ and $q\bar{q}$ lie within the shaded region.

	    This analysis can be easily extended to other resonances, and corrected for geometrical acceptance and efficiency, but the general conclusions remain the same \cite{Chivukula:2016}.

\subsection{Strongly Coupled Dynamics}


Models of strongly coupled dynamics in all their different incarnations \cite{Weinberg:1979bn, Susskind:1978ms, Dimopoulos:1981xc, Kaplan:1983fs, Kaplan:1983sm, Georgi:1984ef, Georgi:1984af, Dugan:1984hq} are among the most popular solutions to the hierarchy problem. In general, in these models, the quadratic ultra-violet (UV) sensitivity of the Higgs mass is saturated at some scale $\Lambda \sim \mathcal{O}(\TeV)$, before the new strong interaction starts to be resolved. In addition, the Higgs boson can be expected to be lighter if it is identified with the pseudo Nambu-Goldstone boson (pNGB) associated to  the spontaneous symmetry breaking  of some global symmetry of the strong sector. This is for instance the case of Composite Higgs models, or their holographic duals, where the Higgs arises as the pNGB of some global symmetry of the strong dynamics~\cite{Contino:2003ve,Agashe:2004rs,Contino:2010rs}, or some walking Technicolor scenarios, where the Higgs is thought to be the dilaton appearing from the breaking of the (approximate) conformal symmetry by the strong dynamics~\cite{Dzhikiya:1986kk,Appelquist:2010gy,Grinstein:2011dq,Matsuzaki:2012gd,Bellazzini:2012vz}.    

The above scale $\Lambda$ is in general  populated by resonances, bound states of UV fields which would couple to the Higgs and massive vector bosons, analogous to the coupling of the QCD $\rho$(770) meson to pions. In the following we will describe proposals to explain the diboson resonance in terms of heavy composite states from such a scenario, exploring different options in terms of quantum numbers, namely $J^{CP}=1^{--}$, $0^{++}$ and $2^{++}$, as well as their couplings to other SM particles besides massive bosons. Note that the production mechanisms for spin-one resonances are via Drell-Yan and (more suppressed) vector boson fusion (VBF), whereas the spin-zero and -two resonances can be produced via gluon fusion. 

\subsubsection{Drell-Yan production: Spin-one resonances}

Typically, in Composite Higgs models one considers the lightest of these states to be a spin-one resonance $\rho_{\mu}$ with vector couplings to the quarks, and in some works a second axi-vector resonance $a_{\mu}$ is also considered. This expectation comes from experience from QCD, as well as dual pictures of Composite Higgs where the SM gauge fields are propagate in the bulk of an extra-dimension. In general, if the UV strong sector would contain light fermions, one would expect these to form bound states with quantum numbers $1^{--}$ which could be at the bottom of the resonance sector.  These spin-one resonances  are expected to feature sizeable couplings with the longitudinal components of the SM weak bosons in analogy to the QCD case, where $g_{\rho\pi\pi}\sim g_{\rho} \gg 1$~\cite{Altmeyer:1995qx}, becoming therefore natural candidates to account for the diboson anomaly observed at the LHC. 

Because of  the aforementioned reasons, it is not surprising that the first models of strongly coupled dynamics trying to address the diboson anomaly that appeared relied on the effect of isospin triplet  vector resonances $\rho^{\pm},\rho^3$ \cite{Fukano:2015hga, Franzosi:2015zra, Thamm:2015csa, Carmona:2015xaa, Bian:2015ota, Lane:2015fza, Low:2015uha, Niehoff:2015iaa}, which are mostly produced by Drell-Yan (DY), 
since the $\mathcal{O}(m_{W}^2/m_{\rho}^2)$ suppression in their couplings to the transverse polarizations of the SM gauge bosons make their VBF production  sub-leading. This is true since, in addition to the $\mathcal{O}(\alpha_W^2)$ suppression of the latter, even in the case of elementary fermions, the coupling to fermions induced by the $\rho-V$ gauge mixing is expected to be $\sim g^2/g_{\rho}$, which is $\mathcal{O}(v^2/f_{\rho}^2)$ bigger than the one involved in VBF, where  $f_{\rho}\sim m_{\rho}/g_{\rho}\gg v$.~\footnote{Note that in the limit of very large $g_{\rho}$, VBF can become relevant. This was explored e.g. in Ref.~\cite{Lane:2015fza}.} Another common feature of these models is the absence of $ZZ$ production, being the decay $\rho^0\to ZZ$ isospin violating. Therefore, the anomaly in such channel can be only explained via leakage from the other channels, $WW$ and $WZ$, which is expected to be large taking into account the ATLAS mass windows defined for the $W/Z$ discrimination. 

From the theoretical point of view, one of the first concerns arising in these scenarios is the problematic interplay of such a \emph{light} resonance with electroweak precision tests (EWPT). The main idea is that tree level $\rho$ contribution to the $S$ parameter will be $\gtrsim 4\pi  v^2/m_{\rho}^2$. Taking $m_{\rho}\sim 2$\,TeV and making numbers, it can be readily seen that if one considers the electroweak fit \cite{Baak:2012kk} at face value, these resonances will be well excluded. However, since they are expected to be part of a more complete theory of electroweak symmetry breaking (EWSB), where new contributions to the oblique parameters are expected, this naive counting has to be taken with a grain of salt.  First of all, in technicolor theories or general composite models where the second Weinberg sum rule can be relaxed \cite{Marzocca:2012zn}, the contribution of the vector resonance to the $S$ parameter can be partially cancelled by the one from its axial counterpart 
\begin{equation}
	S\simeq 4\pi	\frac{v^2}{f_{\rho}^2+f_{a}^2}\left(\frac{f_{\rho}^2}{m_{\rho}^2}-\frac{f_a^2}{m_a^2}\right),
\end{equation}
provided that $f_{\rho}/f_a\sim m_{\rho}/m_a$, where $m_a\sim g_a f_a$. This is actually the case in some technicolor scenarios exhibiting a near conformal behaviour, where nearly degenerate vector and axial resonances are expected \cite{Appelquist:1998xf, Appelquist:1999dq, Hirn:2006nt, Hirn:2006wg, Foadi:2007ue}. Such a cancellation has been 
argued for instance in \cite{Fukano:2015hga, Bian:2015ota, Lane:2015fza}. Moreoever, in the case of \emph{regular} composite Higgs models, the presence of anomalously light top partners is typically required by naturalness~\cite{Contino:2006qr, Medina:2007hz, Csaki:2008zd, Matsedonskyi:2012ym, Marzocca:2012zn, Pomarol:2012qf}, which can lead via  radiative corrections to a more than welcome  positive contribution to the $T$ parameter \cite{Anastasiou:2009rv}, taking into account the strong correlation between both observables.~\footnote{However, their presence  may affect the setups at hand. See discussion below.}   On the other hand, a departure from the universal case (having e.g. different $\rho$ couplings to light quarks and leptons) opens an extra (although limited) room for improvement in the electroweak fit. This is the case e.g. in \cite{Carmona:2015xaa}, where suppressed lepton couplings were required to accommodate dilepton bounds, even though the agreement with EWPT arised mainly from a small breaking of custodial symmetry \cite{Cabrer:2011fb, Carmona:2011ib}. Finally, UV corrections at the cutoff scale may also be important \cite{Contino:2015mha}.

Another potential issue arising in models of composite Higgs, where the fermion masses are generated by partial compositeness,  is that the natural presence of light partners will most likely substantially change the simple picture presented until now. Since these colored fermionic states $Q$ will typically have masses $m_{Q}\ll m_{\rho}\sim g_{\rho}f_{\rho}$, and will feature sizable couplings to the vector resonances $g_{\rho QQ}\sim g_{\rho}$, the diboson branching ratios will be diluted by a much larger total width, if the decay channel $\rho\to Q \bar{Q}$ opens and they can be pair produced \cite{Agashe:2008jb, Barducci:2012kk, Vignaroli:2014bpa, Greco:2014aza}.  Even in the case when only single production of top partners is allowed, since a large degree of compositeness is expected for the top chiralities, the couplings $g_{\rho t Q}$ involved will not be negligible, and their presence will complicate the analyses presented hitherto. However, the presence of anomalously light top partners is not compulsory in natural models \cite{Panico:2012uw, Carmona:2014iwa, Geller:2014kta, Barbieri:2015lqa, Low:2015nqa}  and it is always possible to avoid them at the price of increasing the fine tuning. Alleviating these problems is among the motivations of \cite{Low:2015uha}, where the vector resonances at hand are embedded into a twin Higgs framework that assures the absence of light fermionic partners.  Another related issue is that the sizable degree of compositeness $\epsilon_{L,R}^t$ required to generate the top mass  can make $g_{\rho t_{L,R} t_{L,R}}\sim g_{\rho}\epsilon_{L,R}^t \epsilon_{L,R}^t$ large enough to render the channels $\rho^{0}\to t\bar{t}$ and $\rho^{+}\to t\bar{b}$  competitive with respect to the diboson ones.~\footnote{Smaller mixing angles $\epsilon_{L}^t$ for the third generation are expected in twin Higgs models \cite{Low:2015uha} and also in the UV completion of \cite{Carmona:2015xaa}, even though the required values used in the latter to explain the excess without disagreement with other searches seem a bit on the edge.} 
 To a lesser extent, this is also the case for $\rho^0\to b \bar{b}$, which will also fed into the dijets searches. However, at the end of the day, these constraints turn out to be relevant only for large values of $g_{\rho}$ and $\epsilon_L^t$ \cite{Low:2015uha}, or for moderate values of $g_{\rho}$ and large mixing angles $\epsilon_{L}^q$ \cite{Carmona:2015xaa}, when the dijet branching ratio becomes important. 
 
\begin{figure}[!h]
\centering
\includegraphics[width=0.49\textwidth]{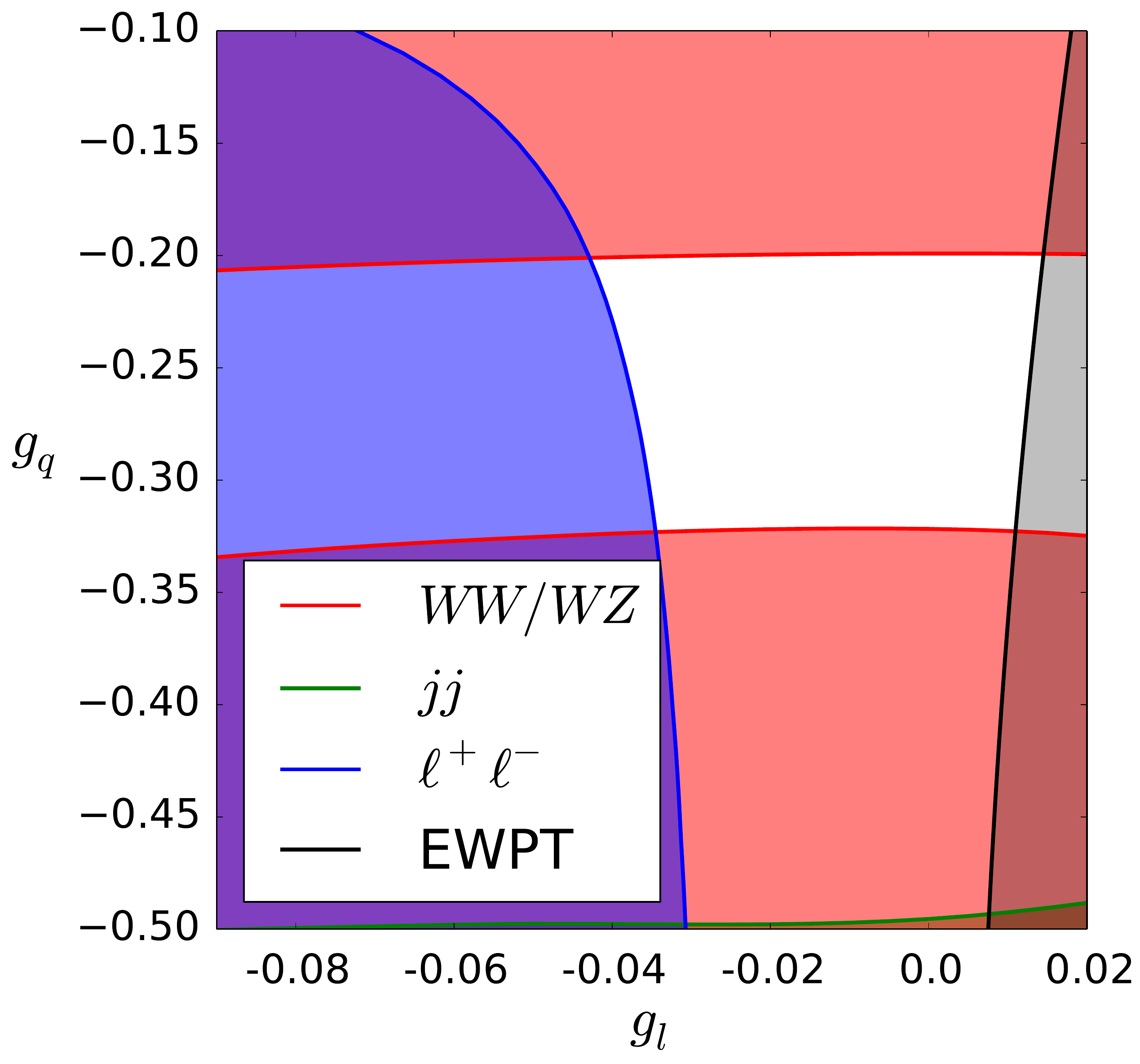}
\includegraphics[width=0.49\textwidth]{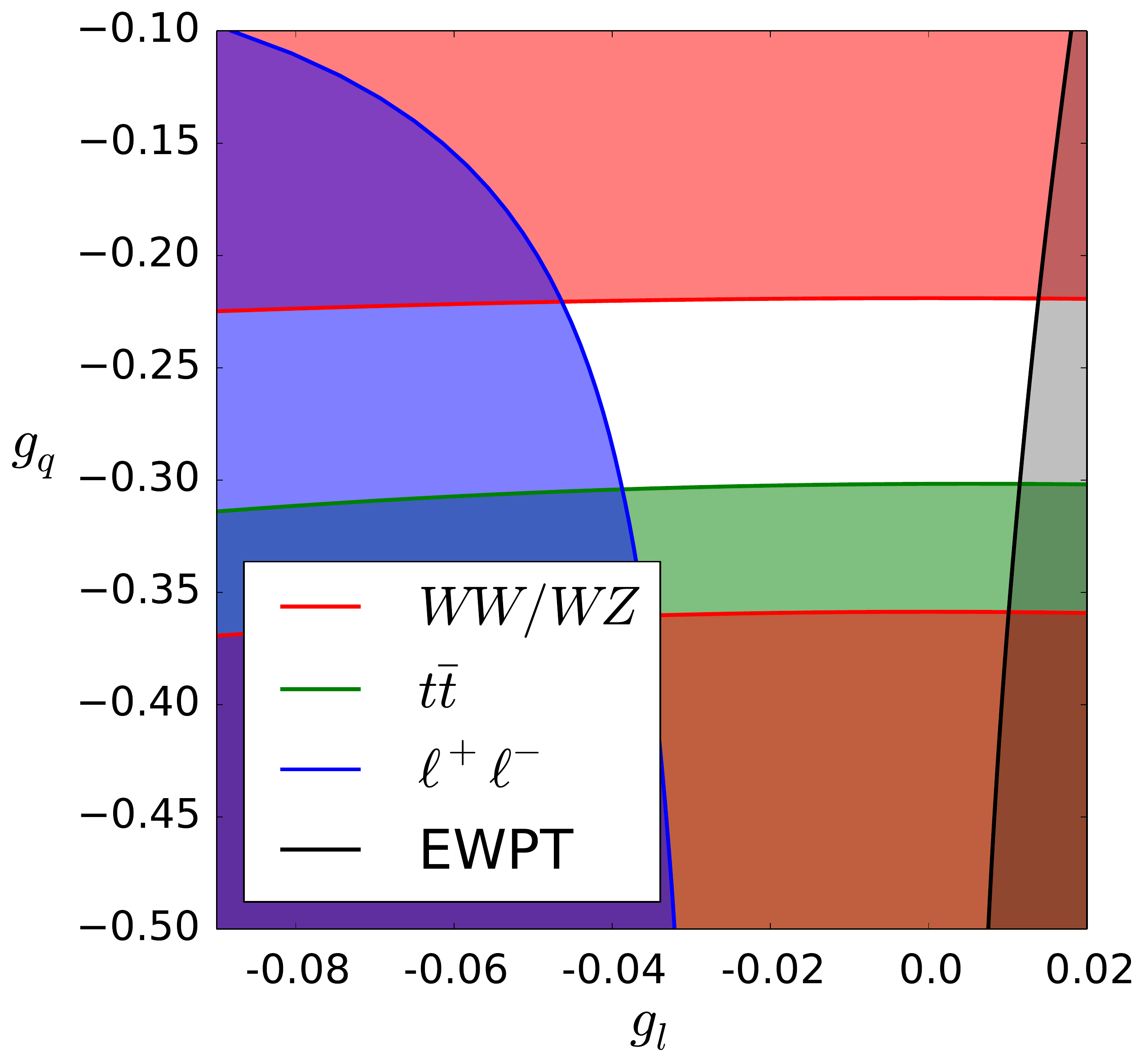}
\caption{Constraints on the  parameter space of the model in Ref.~\cite{Carmona:2015xaa} from the diboson signal, $t\bar{t}$, 
  dijet, $\ell^+ \ell^-$ and EWPT. The colored areas correspond
  to the region excluded by the different bounds.
  Here, $m_{\rho}=1.8$ TeV, $g_{\rho}=0.75$ and $g_{\rho}\epsilon_{L}^{q_3 2}= 0.3$, whereas $g_{\rho}\epsilon_R^{t\, 2}= 0.3\, (0.5)$ for the left (right) plot. See Ref.~\cite{Carmona:2015xaa} for more details.}
\label{fig:smallg}

\end{figure}

\begin{figure}[!h]
\centering
\includegraphics[width=0.49\textwidth]{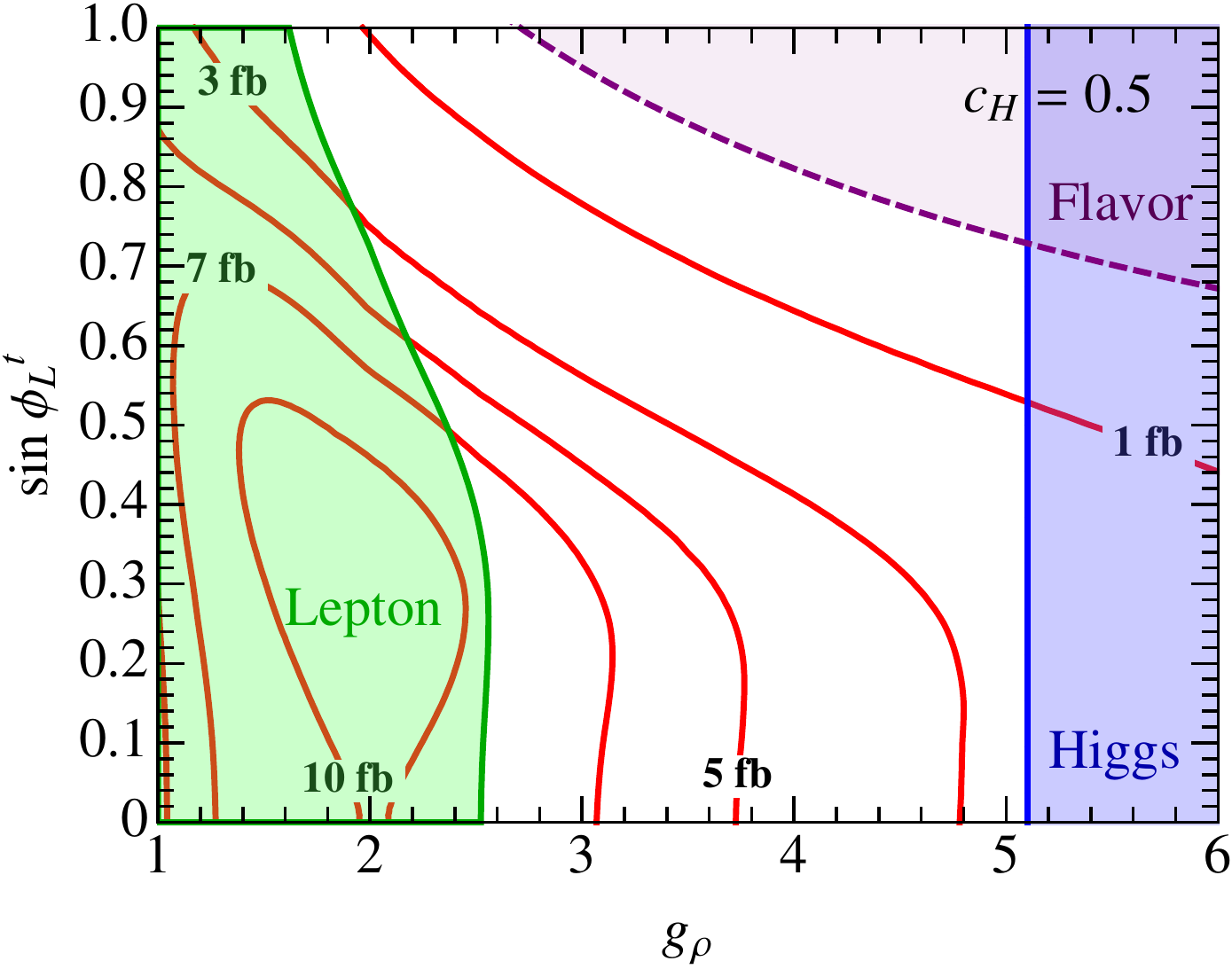}
\includegraphics[width=0.49\textwidth]{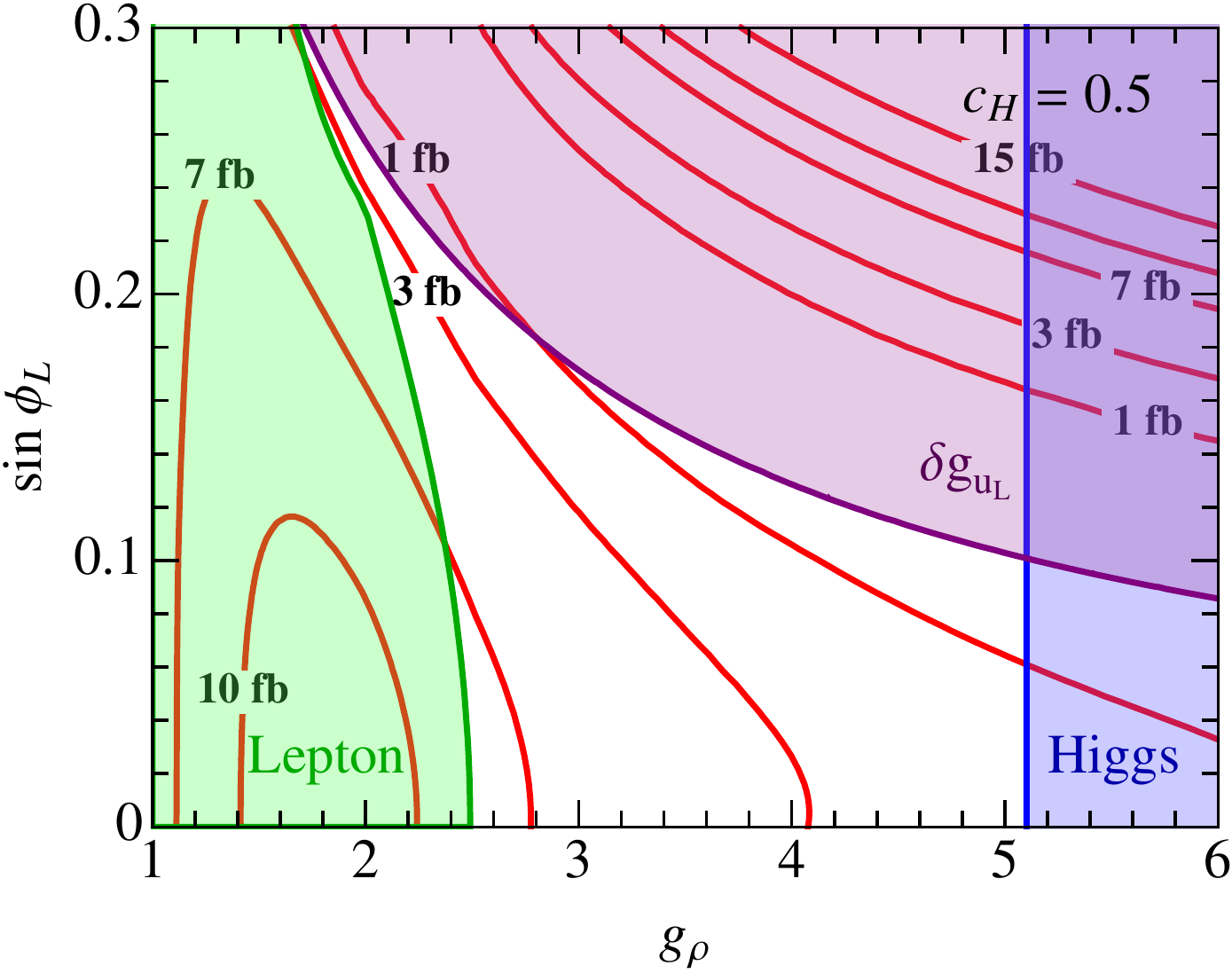}
\caption{Constraints on the parameter space of the two-site model in Ref.~\cite{Low:2015uha} for $m_{\rho}=2\,$TeV. In the left-hand side plot, all leptons and light quarks have been assumed to be elementary and $\sin\phi_L^t\equiv\epsilon_L^t$. In the right-hand side plot, $\sin\phi_L\equiv \epsilon_L^{q_1}\equiv\epsilon_L^{q_2}$, whereas $\epsilon_L^t$ has been fixed to $0.4$. See Ref.~\cite{Low:2015uha} for more details.  
	\label{fig:runningg}
}
\end{figure}

 To illustrate some of these issues and the interplay between the different constraints we show in Figures~\ref{fig:smallg} and \ref{fig:runningg} the viable parameter space for a couple of particular examples, taken from Refs.~\cite{Carmona:2015xaa, Low:2015uha}. In the first case, the relatively small value of $g_{\rho}=0.75$ preferred by EWPT in that model, together with the sizable couplings to light quarks $g_q\sim g_{\rho}(\epsilon_L^q)^2$ required to make  DY production significant, make impossible to neglect  dijet or $t\bar{t}$ searches, as can be readily seen by looking at Figure~\ref{fig:smallg}. However, in the case of models where $T=0$ at the tree level and $g_{\rho}$ is expected to be significantly larger,  like the ones illustrated in Figure~\ref{fig:runningg},  the most relevant direct constraints arise essentially  from dilepton searches when $g/g_{\rho}^2$ starts to become relevant.

 \subsubsection{Gluon-fusion production: Spin-zero and -two resonances}

As we discussed, in the usual Composite Higgs scenarios one needs a baryonic resonance (the top-partner) among the low-lying states to trigger EWSB. This in turn indicates that the UV theory would exhibit a chiral symmetry in analogy to QCD, leading to the expectation that spin-one states composed by light UV quarks would be relatively light. In QCD-like models, as well as in extra-dimensional models, the $1^{--}$ resonances are lighter than the $0^{++}$ and $2^{++}$ bound states, hence their phenomenology is more relevant. 

But this is not the only scenario for a new strong sector in Composite Higgs. An alternative explanation would be that EWSB is triggered by other means not involving top-partners, e.g. via a see-saw mechanism~\cite{Sanz:2015sua}. In this case, the structure of the UV theory can differ from the usual Composite Higgs models, leading to a resonance structure distinctly different from the cases discussed in the previous section.  

\begin{figure}[h!]
\begin{center}
\includegraphics[width=.45\textwidth]{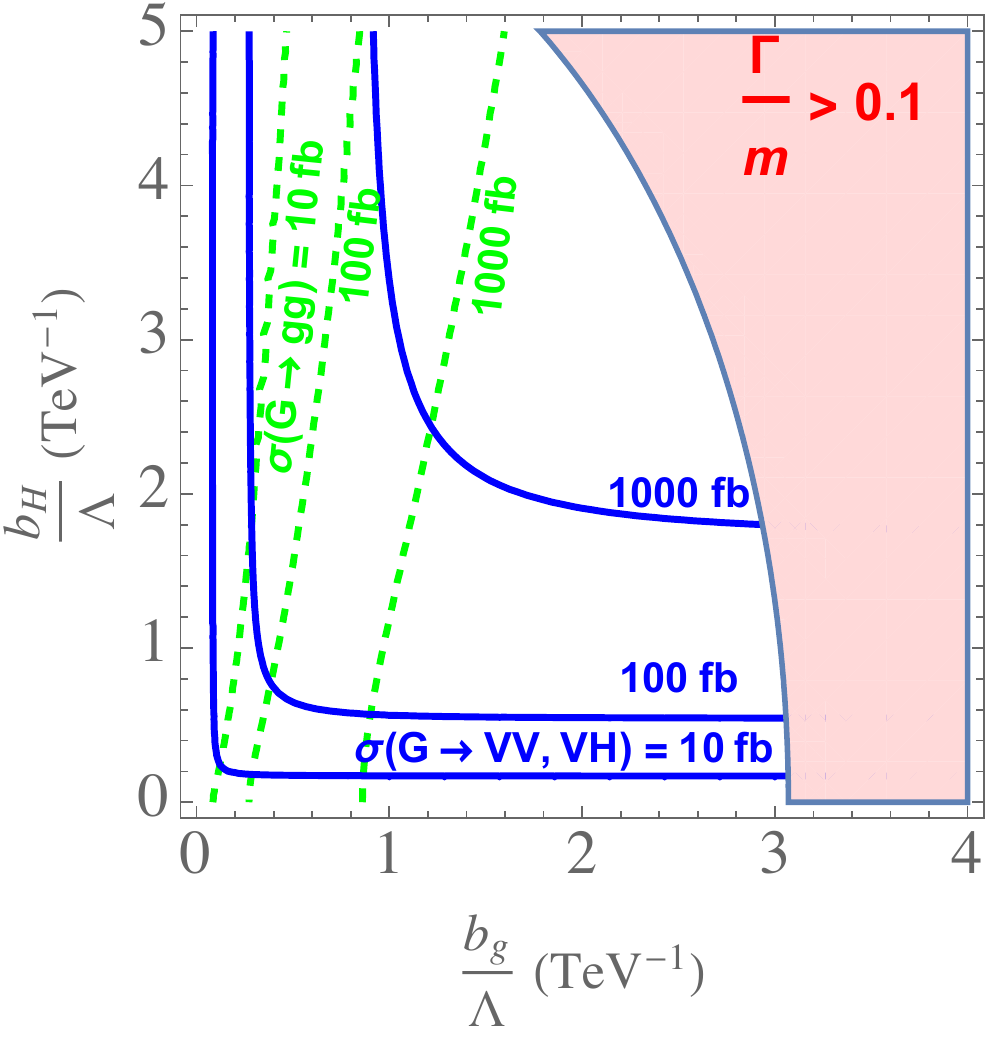}
\caption{Total cross sections at LHC8 for 2 TeV  spin-two states in the $HV$ and $VV$ channels and digluon final state as a function of the coupling of the glueball to gluons and Higgs degrees of freedom. Cross sections do not include efficiencies to cuts, and the shaded region correspond to a width above 10\% of the mass~\cite{Sanz:2015zha}. }
	\label{fig:glueballs}
\end{center}
\end{figure}

Such a scenario in terms of a pure-gauge UV completion has been proposed in Ref.~\cite{Sanz:2015zha}. In this proposal, the lightest states are glueballs, bound states formed by gauge fields. The glueball spectrum has been studied in the lattice~\cite{Teper:1998kw} and one finds that typically the $0^{++}$ and $2^{++}$ states dominate the phenomenology. In~\cite{Sanz:2015zha} it was argued that the behaviour of these glueballs should resemble the radion and Kaluza-Klein graviton states in extra-dimensional theories, with a notable difference that in the glueball theory the spin-one states are no lighter than the massive graviton, hence the strong constraints on the electroweak precision parameters $S$ and $T$ (discussed in the previous section) do not apply in this scenario. 
The glueballs are singlets under the SM interactions and would be produced via gluon fusion, with preferential decays to other (lighter) composite states, namely the Higgs, longitudinal $W$ and $Z$ bosons and possibly the top.  As they are initiated by gluons, the production of these resonances would increase substantially in Run2. See Table~\ref{tab:overview} for a summary of the experimental signatures of this scenario and Fig.~\ref{fig:glueballs} for a plot summarizing the cross section for the distinct spin-two case.

Finally, let us mention that in Ref.~\cite{Chiang:2015lqa}, scalar glueballs were discussed in a scenario with a UV completion involving scalars and unrelated to Composite Higgs, where one would not expect preferential couplings to massive particles. For this reason, the phenomenology of this scenario is different from the glueballs related to Composite Higgs,  as there are no specific predictions of branching ratios.

\subsection{Enlarged Gauge Group}
\label{sec:Gauge}


The decay of a heavy resonance to gauge bosons can arise naturally in
scenarios with an enlarged non-Abelian gauge sector. In such models the
excesses can be interpreted as a new heavy gauge boson decaying
through a triple gauge boson vertex.  A heavy neutral gauge boson is
typically called $Z'$, while a heavy charged gauge boson is usually
denoted by $W'$.

The proposed explanations have some common phenomenological features:
\begin{itemize}
  \item Quarks are charged under the extended gauge groups,
    so the heavy gauge bosons are produced copiously
    via the Drell-Yan process. In fact, the $W'$ or $Z'$ production cross section
    can easily reach a few hundred fb for a mass around 2~TeV and
    weak gauge couplings.
    
  \item This directly implies that the $W'$ or $Z'$ decays into $qq'$,
    leading to potential signatures in dijet, $tb$ and $t\bar{t}$ final states.
    Depending on the model there may also be leptonic decays, but these
    are strongly constrained from $\ell^+ \ell^-$ and $\ell \nu$ searches. 

  \item To explain the diboson excess, couplings to $WZ$, $WW$, or
    $ZZ$ states are necessary. The equivalence theorem \cite{Chanowitz:1985hj} then suggests
    $W' \to Wh$ or $Z' \to Zh$ decay modes of similar size.
\end{itemize}

For concreteness, we will now focus on one particular scenario,
the left-right symmetric model with gauge
group~$SU(2)_L \times SU(2)_R \times U(1)_{B-L}$, in which the excess
is explained in terms of the righthanded $W_R$ boson decaying to $WZ$ pairs.
We will then give an overview of other proposals.

\subsubsection{Left-right symmetry}

In the context of the diboson excess at the LHC, the most studied framework
with an enlarged gauge sector is the left-right symmetric model based on the gauge
group $SU(2)_L \times SU(2)_R \times U(1)_{B-L}$~\cite{Pati:1974yy,
  Mohapatra:1974hk, Mohapatra:1974gc, Senjanovic:1975rk, Mohapatra:1986uf,
Deshpande:1990ip}.
Recent studies in the context of the diboson anomaly have been presented in
Refs.~\cite{Dhuria:2015cfa,Hisano:2015gna,
Cheung:2015nha, Dobrescu:2015qna, Gao:2015irw,
Brehmer:2015cia, Cao:2015lia, Heeck:2015qra, Allanach:2015hba,
Dobrescu:2015yba,Dev:2015pga, Coloma:2015una,
Deppisch:2015cua,
Bandyopadhyay:2015fka, Awasthi:2015ota, Ko:2015uma,
Collins:2015wua, Dobrescu:2015jvn,
Das:2015ysz,Aguilar-Saavedra:2015yza,
Bambhaniya:2015ipg,Hirsch:2015fvq,
Aydemir:2015oob} (see also Ref.~\cite{Bian:2015hda} for a study
utilizing a $W'$ toy model, which can be considered a phenomenological
reduction of the full left-right symmetric model).
In models of this type, right-handed fermions are collated
into doublets of the new gauge group $SU(2)_R$ in the same way as their
left-handed counterparts are members of $SU(2)_L$ doublets. The model can thus
restore parity as an exact (but spontaneously broken) symmetry of nature.

\begin{figure}
  \centering
  \includegraphics[width=0.7\textwidth]{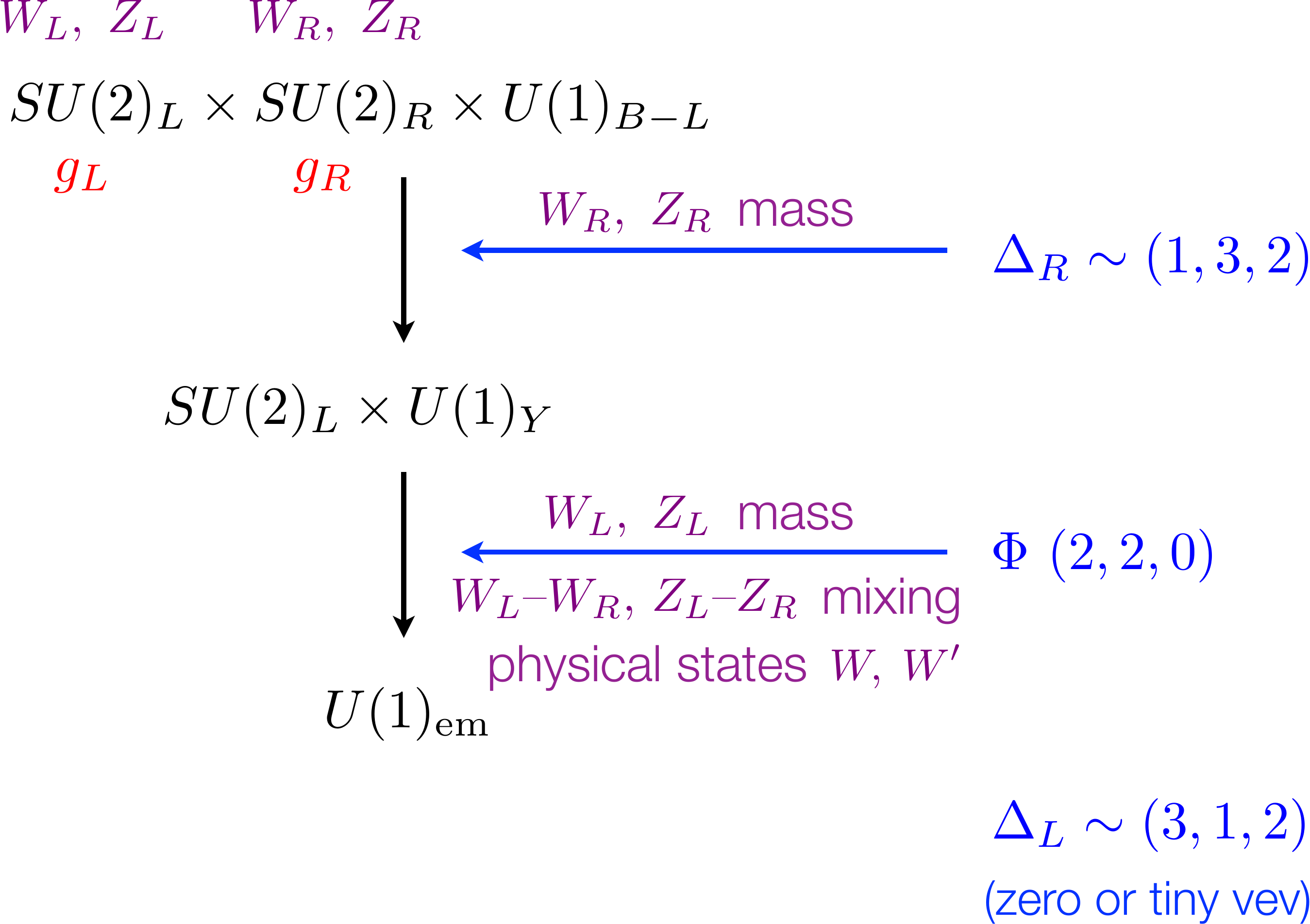}
  \caption{A typical breaking scheme for the left-right symmetric gauge group
    $SU(2)_L \times SU(2)_R \times U(1)_{B-L}$.}
  \label{fig:LR-breaking}
\end{figure}

At a scale~$\gg 100$~GeV, $SU(2)_R \times U(1)_{B-L}$ is spontaneously broken
to $U(1)_Y$, with the most popular breaking mechanism involving a triplet Higgs
field $\Delta_R$ with quantum numbers $(1,3,2)$ under $SU(2)_L \times SU(2)_R
\times U(1)_{B-L}$ (see Fig.~\ref{fig:LR-breaking}).  The {\it vev} of $\Delta_R$
gives mass to the $SU(2)_R$ gauge bosons $W_R$, $Z_R$.  At the electroweak
scale, a Higgs bidoublet $\Phi \sim (2, 2, 0)$ then breaks the residual
$SU(2)_L \times U(1)_Y$ to $U(1)_\text{em}$, also endows the $SU(2)_L$ gauge
bosons $W_L$, $Z_L$ with masses, and leads to a small mixing between $W_L$,
$W_R$ as well as between $Z_L$, $Z_R$.  We thus obtain two physical charged gauge
bosons $W \simeq W_L$ and $W' \simeq W_R$, with $m_W \ll m_{W'}$. Similarly, the
physical neutral gauge boson states $Z$ and $Z'$ satisfy $m_Z \ll m_{Z'}$.
This breaking scheme of the left-right symmetric gauge group has the additional
benefit that the vev of $\Delta_R$ generates a Majorana mass term for the right
handed neutrinos (which are automatically part of the model thanks to the
doublet structure of right handed fermions).

At the LHC, $W'$ bosons can be efficiently produced through their coupling
to right-handed quarks. They subsequently decay either to quarks (leading to
dijet and $tb$ resonance signatures), via $W' \to W Z$ (explaining
the diboson signature), or into $Wh$. The latter decay modes are generated by the
small mixing between $W_L$ and $W_R$. Depending on the mass of the right-handed
neutrinos, there may also be a $W' \to \ell N_R$ decay~\cite{Dhuria:2015cfa,
Dobrescu:2015qna,Dobrescu:2015yba,Coloma:2015una,
Deppisch:2015cua,Awasthi:2015ota,Ko:2015uma,Collins:2015wua,
Dobrescu:2015jvn,Bambhaniya:2015ipg,Hirsch:2015fvq}.
$W'$ decays into the additional Higgs bosons of
the extended scalar sector~\cite{Dobrescu:2015yba}, or into new fermion
multiplets~\cite{Brehmer:2015cia,Heeck:2015qra,Dev:2015pga,
Ko:2015uma,Collins:2015wua} have also been discussed.
Finally, the authors of Ref.~\cite{Aguilar-Saavedra:2015iew} have analysed
three-body and four-body decays.

\begin{figure}
  \centering
  \includegraphics[width=0.49\textwidth]{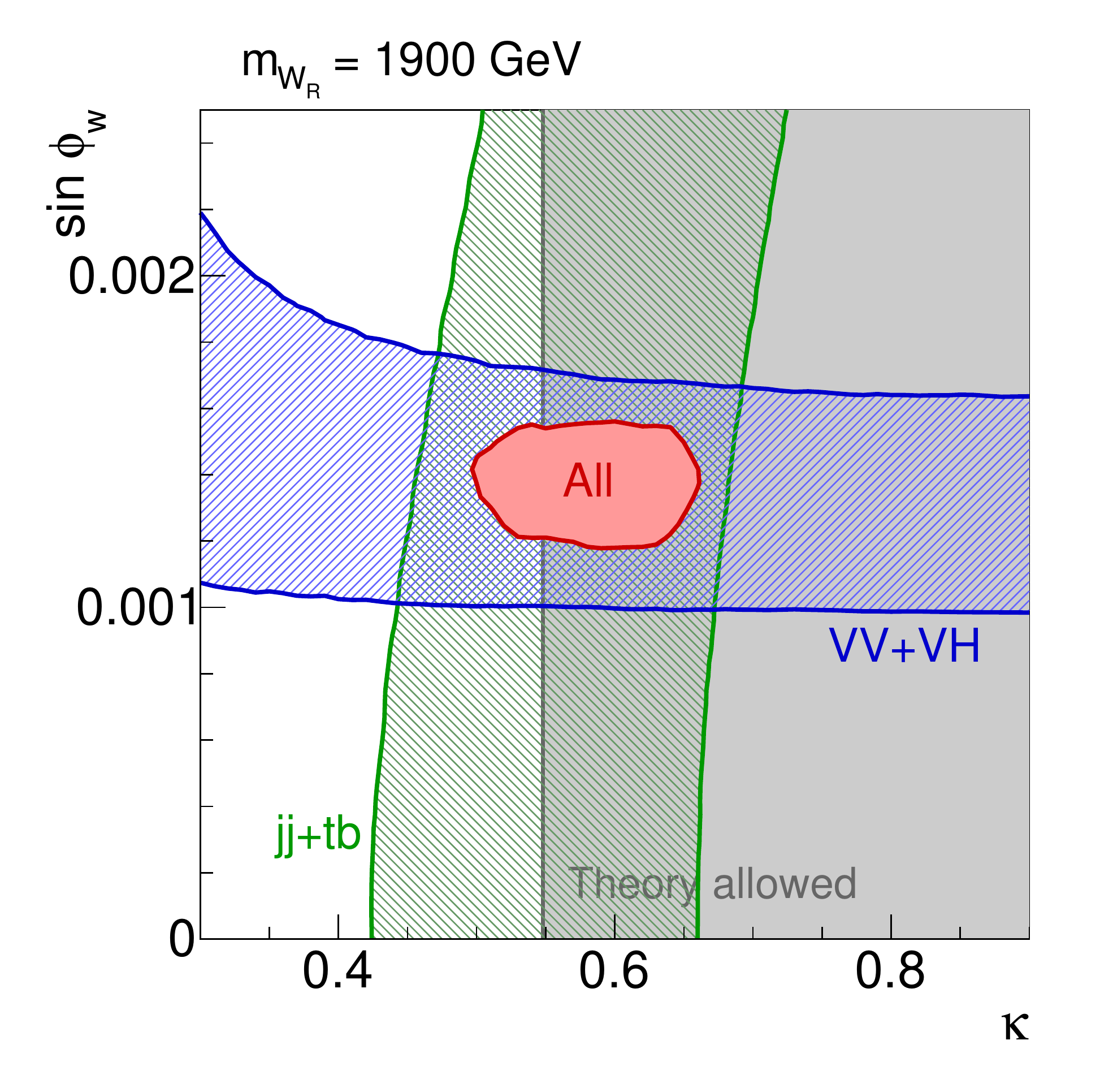}
  \includegraphics[width=0.49\textwidth]{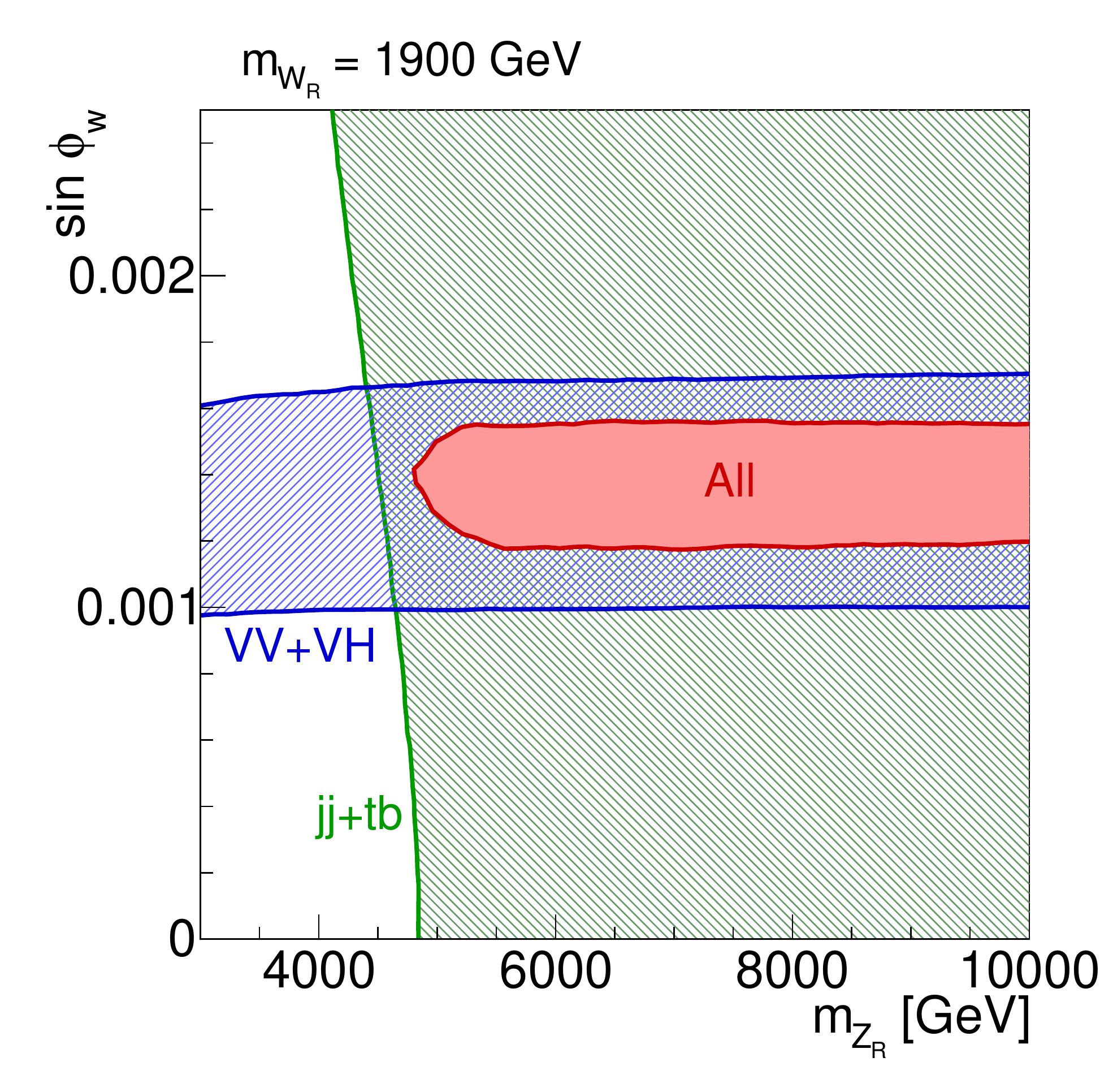}
  \caption{Preferred parameter region in the left-right symmetric model.
    This explanation of the diboson excess depends on only three
    parameters: the $W'$ mass $m_{W'} \sim m_{W_R}$, the ratio
    $\kappa = g_R / g_L$ of the $SU(2)_L$ and $SU(2)_R$ gauge couplings,
    and the $W_L$-$W_R$ mixing angle $\phi_w$.
    In terms of these parameters, a lower limit
    on the $Z_R$ mass can be derived (on the right).
    The blue (green) bands show the parameter regions in agreement with the
    diboson excess and $Vh$ constraints (dijet and $tb$ searches) at $68\%$~CL.
    The region favoured by all searches combined is shown in red.
    Coupling constant ratios $\kappa < 0.55$ are forbidden by
    the condition that the breaking of $SU(2)_R \times U(1)_{B-L}$ to
    $U(1)_Y$ must yield the correct value for the hypercharge coupling
    constant. Figure taken from Ref.~\cite{Brehmer:2015cia}.}
  \label{fig:LR-fit}
\end{figure}

It turns out that this right-handed $W'$ boson can explain the diboson
excess while being consistent with all other current limits from LHC searches.
For good agreement with data, the following conditions must be
satisfied (see Fig.~\ref{fig:LR-fit}):
\begin{itemize}
  \item The $W'$ boson must have a mass $m_{W'} \sim 1.8 \dots 2.0$~TeV.

  \item The $SU(2)_R$ coupling constant $g_R$ must be smaller than the
    $SU(2)_L$ coupling constant $g_L$ to avoid a too large $pp \to W' \to qq'$
    cross section in conflict with limits from dijet and $tb$ searches.
    In particular, the fit from Ref.~\cite{Brehmer:2015cia} finds
    $\kappa \equiv g_R / g_L \sim 0.55 \dots 0.7$ as the preferred region
    for $m_{W'} \sim 1.9$~TeV.
    Note that values $\kappa < 0.55$, while experimentally allowed,
    lead to a theoretical inconsistency in the model. At such small $g_R$,
    the model cannot reproduce the correct $U(1)_Y$ coupling $g_Y = (1/g_R^2 +
    1/g_{B-L}^2)^{-1/2}$ after $SU(2)_R \times U(1)_{B-L}$ breaking.
    Of course, a scenario with $g_L \neq g_R$ may seem less aesthetic
    than one with $g_L = g_R$, but a difference between
    the two coupling constants is expected from renormalization group evolution \cite{Rizzo:1981jr,Chang:1983fu}
    and therefore does not prevent $SU(2)_L \times SU(2)_R \times U(1)_{B-L}$
    from unifying into a larger gauge group at a higher scale.

  \item The $W_L$-$W_R$ mixing angle should be of order $\sin\phi_w \sim
    0.001 \dots 0.002$ to give the correct $W' \to WZ$ cross
    section~\cite{Brehmer:2015cia}.
\end{itemize}

These three parameters $m_{W'}$, $\kappa$, and $\phi_w$ not only
determine the $W'$ signal cross sections in the $WZ$, $Wh$,
dijet, and $tb$ final states, but also fix the
$Z'$ mass. Based on the experimental data, $m_{Z'}$ should be $\gtrsim \text{few TeV}$
(see the right panel of Fig.~\ref{fig:LR-fit}). Values much above
the threshold are experimentally allowed, but require severe
tuning of $\kappa$ close to the theoretical limit 0.55.

It is worth mentioning that a left-right symmetric interpretation of the
diboson excess can provide a few additional features:
\begin{itemize}
  \item The left-right symmetric model might also allow an explanation of the
    CMS anomaly  in the $ee jj$ final state~\cite{Khachatryan:2014dka} with a
    $2.8\,\sigma$ significance.
    This excess can be interpreted as a signal of the decay $W' \to e N_R
    \to e e + (W'^* \to jj)$, where $N_R$ is a heavy right-handed neutrino.
    Since CMS observe an excess only in opposite-sign electrons, while the data
    in the same-sign electron (and in the $\mu\mu j j$) channels are consistent with
    SM predictions, Refs.~\cite{Dobrescu:2015yba,Dobrescu:2015qna,Deppisch:2015cua,
    Dobrescu:2015jvn} propose that at least one of the $N_R$ is a pseudo-Dirac
    neutrino,  thus suppressing the $e^\pm e^\pm j j$ final state compared to the
    $e^\pm e^\mp j j$ one.

  \item The model can be embedded in larger gauge groups such as
    an $SO(10)$ GUT~\cite{Deppisch:2015cua,Bambhaniya:2015ipg,Aydemir:2015oob}
    or in a non-commutative geometry framework~\cite{Aydemir:2015nfa}.

  \item Possible connections between the left-right symmetric model and dark
    matter (DM) have been explored in the
    literature~\cite{Brehmer:2015cia,Heeck:2015qra}. These include
    (1)~$W'$-mediated couplings between a right-handed neutrino DM candidate $N$ 
    and an SM charged lepton. This scenario can only work if $N$ is lighter
    than the charged lepton it couples to, and if mixing with other
    right-handed neutrinos is absent to prevent DM decay. (2)~A new
    $SU(2)_R$ multiplet. In this case, the DM candidate is accompanied by
    one or several new charged states potentially observable at the LHC.
    This scenario has the additional feature that the branching ratio of
    the $W'$ boson to visible final states may be reduced, thus allowing for larger
    values of $\kappa$. (3)~Supersymmetric dark matter. In supersymmetric
    left-right models, the lightest superpartner is an excellent DM
    candidate, in analogy to the MSSM.
\end{itemize}

\subsubsection{Other models with extended gauge groups}

While left-right symmetric scenarios have received the most attention in the
context of the diboson anomaly, other classes of models are equally attractive.
We begin by considering further scenarios in which the diboson anomaly is explained by
a new charged gauge boson $W'$ decaying to a $WZ$ final state, or by a combination of
approximately mass-degenerate $W'$ and $Z'$ bosons.

First, Refs.~\cite{Cheung:2015nha,Allanach:2015hba} consider phenomenological
models with a $W'$ and a $Z'$ boson, for instance based on the gauge group
$SU(2)_L \times SU(2)_R \times U(1)_X$.
Unlike in the scenarios discussed in the previous section, here the $W'$ and $Z'$ bosons
can have equal masses and share responsibility for the observed excesses. 

It is worth noting that ``221 models'' (models with gauge group $SU(2)
\times SU(2) \times U(1)$) can also be interesting in contexts different from
left-right symmetry~\cite{Cao:2015lia,Abe:2015jra,Abe:2015uaa}. 
For instance, different assignments of fermion quantum numbers can lead
to scenarios in which the $W'$ boson couples
in a leptophobic, hadrophobic, fermiophobic, or flavor non-universal way.  The
authors of Ref.~\cite{Abe:2015jra} consider the gauge group $SU(2) \times SU(2)
\times U(1)$ in the context of a 3-site moose model in a deconstructed extra
dimension, analysing both the scenario where only the $W'$ boson is responsible
for the diboson excess as well as the case with nearly mass-degenerate $W'$ and
$Z'$ bosons around 2~TeV.

To supplement their study of 221 models, the authors of Ref.~\cite{Cao:2015lia}
have also considered a 331 model, i.\,e.\ a model in which the gauge group
$SU(3)_c \times SU(3)_L \times U(1)_X$ is broken to the SM gauge
group~\cite{Pisano:1991ee,Foot:1992rh,Frampton:1992wt}.  This leads to the
emergence of five heavy gauge bosons: two pairs of charged bosons and one extra
neutral boson that might be considered for the diboson excess. However, it
turns out that this explanation is not favored by the data~\cite{Cao:2015lia}.

Very recently, the authors of Ref.~\cite{Appelquist:2015vdl} have considered a
model with symmetry group $[SU(2)]^4 \times U(1)_{B-L}$, which preserves not only
left-right symmetry, but also custodial symmetry. This effective theory
is mainly motivated by strongly interacting models, but can also represent
a weakly interacting extended gauge sector. It allows for an explanation of the excess
as a combination of two charged and two neutral gauge bosons at nearly
the same mass, with the majority of the observed events coming from one of the
charged bosons decaying as $W' \to WZ$ (see also Ref.~\cite{Lane:2015fza}).

Since a clear discrimination between $WZ$, $WW$, and $ZZ$ final states is
currently not possible in a boosted diboson search, the ATLAS excess can also
be explained in models with a $Z'$ boson decaying to $WW$ or $ZZ$.
There is a sublety, though: it is difficult to achieve a
sizable $Z'$ coupling to $ZZ$ states in simple models.
On the other hand, an explanation of the ATLAS excess purely in terms
of the decay $Z' \to (W \to jj) + (W \to jj) $ is in tension with strong
constraints from searches for semileptonic final states,
i.\,e.\ $Z' \to (W \to \ell\nu) + (W \to j j)$. 
Nevertheless, a plethora of explanations in terms of a $Z'$ boson from an extra $U(1)$
group has been put forward~\cite{Hisano:2015gna,Alves:2015mua,
Anchordoqui:2015uea,Faraggi:2015iaa,Li:2015yya,Wang:2015sxe,
Feng:2015rzn}. In most of these models the $Z'$ boson is leptophobic to avoid
the strong limits from $\ell^+ \ell^-$ searches~\cite{Hisano:2015gna,
Faraggi:2015iaa,Li:2015yya,Wang:2015sxe,Feng:2015rzn}.

Refs.~\cite{Anchordoqui:2015uea,Faraggi:2015iaa,Li:2015yya}
consider $U(1)'$ extensions of the SM motivated by string theory.
The heavy $Z'$ boson associated with this extra
gauge factor can again be produced through a coupling to light quarks. 
In the model presented in Ref.~\cite{Anchordoqui:2015uea},
the cancellation of anomalies through the Green-Schwarz mechanism
leads to a massive $Z'$ boson with an effective coupling to two electroweak gauge bosons.
This boson then decays not only into $WW$, but also to
$ZZ$ pairs, and can thus explain the diboson excess without
violating semileptonic $WW$ limits.
At the same time, the effective coupling provides a $Z' \to Z \gamma$
decay. Such a signal is one of the hallmark predictions of this class of models.

\subsection{Extended Higgs Sectors}
\label{sec:Scalar}


There have been several attempts ({\it e.g.}~\cite{Chen:2015xql, Omura:2015nwa, Chao:2015eea}) 
to explain the observed diboson excess(es) 
as produced by $\sim 2$ TeV scalars originating in an extended Higgs sector, like a Two Higgs Doublet Model (2HDM). 
Here we discuss the generic features of such scenarios, and give a detailed overview of the problems they face in explaining the diboson excess. Before we go in the details of the models let us just emphasize several common features 
of these types of scenarios. 
\begin{itemize}
\item The entire excess is coming from a neutral heavy Higgs boson, which is produced abundantly at the LHC and decays 
into $ZZ$ and $WW$
\item In order to have sufficient production cross sections, big couplings to the light quarks are introduced.  Thus
these scenarios are necessarily very far from MFV scenarios and they demand extremely severe fine-tuning 
in order to avoid abundant Falvor Changing Neutral Currents (FCNCs) in essentially almost all flavor measurements. 
\item In order not to be in direct conflict with the Higgs couplings measurements, the 2HDM must be in near 
alignment limit, which is in tension with getting large enough BRs into $WW$ and $ZZ$.   
\end{itemize}


Let us at this point focus for concreteness on a 2HDM scenario. 
This features two CP-even neutral scalars, $h$ and $H$, the former being the $125$ GeV Higgs particle while the latter
is considered as the $2$~TeV particle responsible for the diboson excess via its decays $H \to WW, ZZ$. 
In order for $H$ to have a sufficiently large production cross
section, it needs to couple sizably to the first-generation quarks. On the other hand, 
a large coupling of the heavy Higgs boson to the first generation fermions will produce 
a large mass for the up or down quark, which would be unacceptable. In order to avoid this 
problem it is assumed that one doublet {\it does not develop a vev}, namely one 
parametrizes the Higgs doublets as follow:
 %
 \begin{equation}
H_1 = \left(\begin{array}{c}
      G^{+} \\
\frac{v + \varphi_1 + i G_0 }{\sqrt{2}}
      \end{array}\right)
      \quad \quad \quad 
H_2 = \left(\begin{array}{c}
       H^{+} \\
\frac{\varphi_2 + i A_0 }{\sqrt{2}}
      \end{array}\right)      
\end{equation}
with $H^{+}$ and $A_0$ being respectively the physical charged and neutral CP-odd mass eigenstates, and $G^{+}$, $G_0$ being the Goldstone bosons.  Namely we are in peculiar situation where $\tan \beta = \infty$, however 
the mixing angle between the CP even states $\alpha$ is different from $\pi/2 $ or $0$. This is in principle 
possible to achieve by carefully balancing the coefficients $m_{12}^2$ and $\lambda_6$ 
using the conventions of Ref.~\cite{Gunion:2002zf}. However, this situation is highly non-generic and also
demands non-trivial fine-tuning.

In the \textsl{alignment limit} ($\mathrm{cos}(\beta - \alpha) = 0$) of the 2HDM, $H_1 \equiv H_{\mathrm{SM}}$, and then $\varphi_1 = h$ is the SM Higgs boson. Away from this limit, 
the CP-even states $\varphi_1$ and $\varphi_2$ mix into the mass eigenstates $h = \mathrm{sin}(\beta - \alpha) \varphi_1 + \mathrm{cos}(\beta - \alpha) \varphi_2$, 
$H = -\mathrm{cos}(\beta - \alpha) \varphi_1 + \mathrm{sin}(\beta - \alpha) \varphi_2$. It is through this mixing that $H$ can decay to $WW$ and $ZZ$.
Turning now to the Yukawa Lagrangian for the two scalar doublets $H_{1,2}$: 
 \begin{equation}
 \label{Yukawas}
\mathcal{L}_Y = - \overline{Q}^i_L H_1\, y^i_d d^i_R - \overline{Q}^i_L (V^{\dagger}_{\mathrm{CKM}})^{ij}\tilde{H}_1\, y^j_u u^j_R 
- \overline{Q}^i_L H_2\, Y^{ij}_d d^j_R - \overline{Q}^i_L (V^{\dagger}_{\mathrm{CKM}})^{ij}\tilde{H}_2\, Y^{jk}_u u^k_R 
 \end{equation}
where $Q = (V^{\dagger}_{\mathrm{CKM}}u_L, d_L)^T$ and the quarks are already written in the mass eigenbasis. 
The rationale in Refs.~\cite{Chen:2015xql, Omura:2015nwa, Chao:2015eea} is that, as $H_2$ 
does not develop a {\it vev}, and thus does not contribute to 
the quark masses, the couplings of $H_2$ to the first generation of quarks may be large, 
{\it e.g.} $Y^{11}_u = \sqrt{2}m_u/v \times \xi_u$, with $\xi_u \gg 1$.
We however stress that generically $Y^{ij}_{u,d}$ in Eq.~\eqref{Yukawas} would be $3 \times 3$ complex matrices 
with off-diagonal entries, 
leading to FCNCs
mediated by $H$, and by virtue of the CP-even mixing, also by $h$. Even if $Y^{ij}_{u,d}$ would be (ad-hoc) 
taken to be diagonal at tree-level, this choice is not 
protected by any symmetry and would not be preserved by renormalization group evolution. 
This issue poses a serious threat to these models, but has not been analyzed 
in Refs.~\cite{Chen:2015xql, Omura:2015nwa, Chao:2015eea}.


We now analyze the viability of such scenarios as an explanation of the diboson excess. 
First, due to the CP-even mixing, the light Higgs $h$
will inherit the large $H_2$ couplings to first-generation quarks in~\eqref{Yukawas}, 
which may affect the Higgs signal strengths through an increase in Higgs production.
The production cross section in $u \bar{u}$ initial states for $h$ and $H$ (at LO) is 
\begin{eqnarray}\label{eq:ExoticHProd}
\sigma (p p\, (u \bar u) \to h) \sim \left( \sqrt{2}\,\mathrm{cos}(\beta-\alpha)\, \xi_u\,m_u/v\right)^2 \times 54\,\,\, \mathrm{nb}, \\
\sigma (p p\, (u \bar u) \to H) \sim \left( \sqrt{2}\,\mathrm{sin}(\beta-\alpha)\, \xi_u\,m_u/v\right)^2 \times 0.52\,\,\, \mathrm{pb}. 
\end{eqnarray}
For $m_u \sim 2.3$ MeV,  $\xi_u \sim 10^4$ and $\mathrm{cos}(\beta-\alpha) \sim 0.01-0.1$ (as considered in Refs.~\cite{Omura:2015nwa, Chao:2015eea}), 
$\sigma (p p \,(u \bar u) \to h) \sim  0.1-10\,\,\, \mathrm{pb}$, to be compared to $\sigma (p p\, (g g) \to h) \sim 19.17\,\,\, \mathrm{pb}$ in the SM (computed however at
NNLO QCD and NLO EW). This sets an upper bound on 
the mixing parameter $\mathrm{cos}(\beta-\alpha)$ (as a function of $\xi_u$), and constitutes an important constraint on 
these models. 


The necessary cross section to fit the diboson excess is $\sigma^H_{VV} \equiv \sigma(p p \to H) \times \mathrm{BR}(H \to WW,ZZ) \gtrsim 5$ fb. 
The relevant partial decays widths of $H$ are  
\begin{equation}
\Gamma_{H\to jj} \sim \frac{6\,m^2_u\,\xi^2_u\,m_H}{8 \pi v^2} \quad \quad \Gamma_{H\to VV} \sim \frac{m^3_H}{16 \pi v^2} \mathrm{cos}^2(\beta-\alpha)
 \quad \quad \Gamma_{H\to hh} \sim 9\, \Gamma_{H\to VV}.
\end{equation}
The last relation follows from the equivalence theorem (see {\it e.g.} Ref.~\cite{Katz:2014mba}, and for 
more precise numerical estimates see Ref.~\cite{Craig:2013hca}). The decay $H\to hh$ has however not been included in the 
analyses
of \cite{Omura:2015nwa, Chao:2015eea}, and its inclusion will significantly affect the possibility of achieving the required $\sigma^H_{VV}$, particularly 
if $\Gamma_{H\to VV} \gg \Gamma_{H\to jj}$. As an example, Ref.~\cite{Omura:2015nwa} states that $\mathrm{cos}(\beta-\alpha) = 0.05$, $\xi_u = 0.8 \times 10^4$ 
yield $\sigma^H_{VV} \sim 5$ fb, while with the inclusion of $H\to hh$ and for fixed $\mathrm{cos}(\beta-\alpha) = 0.05$, the needed value for $\sigma^H_{VV} \sim 5$ fb 
is $\xi_u = 3.6 \times 10^4$, very close to the exclusion limit from dijet searches.



Another constraint to worry about
has to do 
with the observed production cross sections and BRs of the observed Higgs boson. First of all, because 
in this model there are two states, which unitarize the $WW$ scattering, the coupling of the SM-like
Higgs boson to the vector bosons is modified by $r_V = \sin (\beta -\alpha) $. For $\cos (\beta - \alpha ) \lesssim 0.1$
this leads to a very minor deviation of less than 1\%. Similar deviations of $\sin (\beta - \alpha)$ are expected 
in couplings to the fermions if   the matrices $Y_u$ and $Y_d$ are assumed to vanish. However, to have a 
large coupling between the heavy Higgs boson and the up-type quark we should at least assume that $(Y_u)^{uu} \sim 0.1$,
triggering the new Higgs production mode according to Eq.~\eqref{eq:ExoticHProd}. This coupling is 
virtually indistinguishable from the gluon couplings, and in the near alignment limit this is the only noticeable deviation 
of the Higgs couplings from the SM predictions. After LHC8 the measured deviation of the $hgg$ coupling from the SM value 
is approximately $r_g \approx 0.87 \pm 0.2$ \cite{Giardino:2013bma},
 such that deviations of order 30\% are still allowed.   



We finally comment on other possible spin-0 resonances as being responsible for the diboson excess.
In Ref.~\cite{Sajjad:2015urz} it has been demonstrated, concerning a diboson excess 
originating in the $WZ$ final state, that $SU(2)_L$ singlet, charged spin-0 states cannot be responsible. 
Even though the arguments do not strictly apply to $H^{+}$ from the 2HDM,
in this case the conclusions are similar as there exist no tree-level coupling $H^{\pm} W^{\mp}Z$.


\section{Phenomenology for Run 2}
\label{sec:Pheno}

We present here a discussion of how LHC Run 2 might determine the general properties of the diboson excesses near 2 TeV, under the assumption that they arise from fundamental (gauge or Higgs bosons) or composite states of new beyond the standard model intearctions.  First we make some general remarks about production and decay modes of resonances from each of the categories of models discussed in section 3.  Then we propose several specific measurements that can help pin down the character of these interactions.

\vspace{2mm}
\subsection{Resonance Production and Decay Modes}

As discussed in Sect.~3, theoretical proposals for new physics responsible
for the ATLAS and CMS 2~TeV excesses fall into a few broad categories:

\begin{itemize}

\item[1)] New strong dynamics, involving heavy $\rho$ and, possibly,
  $a_1$-like vector bosons associated with the 125~GeV Higgs boson $H$ being
  a composite structure of some sort, e.g., Refs.~\cite{Franzosi:2015zra,
    Fukano:2015hga,Carmona:2015xaa,Thamm:2015csa,Lane:2015fza,
    Appelquist:2015vdl}.

\item[2)] Extended gauge dynamics, generally involving weakly-coupled heavy
  $W'$ and/or $Z'$ bosons; e.g., Refs.~\cite{Cheung:2015nha,Dobrescu:2015qna,
    Brehmer:2015cia}.

\item[3)] Extended Higgs-sector models, also generally involving weak
  coupling; see, e.g., Refs.~\cite{Chen:2015xql,Omura:2015nwa,
    Chao:2015eea}. Also see Refs.~\cite{Chiang:2015lqa, Cacciapaglia:2015nga}
  for spin-zero models inspired by strong dynamics.

\item[4)] Spin-two or effective operator frameworks that consider multiple
  spins.

\end{itemize}

The principal production mechanism at the LHC of the vector bosons of new
strong or weak dynamics is the Drell-Yan (DY) process of $\bar qq'$
annihilation. For 2~TeV masses, their production rate at 13~TeV will be 5-7
times greater than at 8~TeV. In new strong dynamics, the vectors typically do
not couple directly to standard model (SM) quarks and leptons, so their DY
production proceeds via their mixing with $\gamma,W,Z$ bosons. New weakly-coupled
gauge bosons do couple directly to quarks and leptons. A secondary, but
non-negligible production mechanism for these vector bosons, especially those
associated with strong dynamics, is weak vector boson fusion (VBF), usually
involving longitudinally-polarized $W_L, Z_L$ bosons. The VBF rate of the vectors
increases by an order of magnitude at 13~TeV. In the case of extended
Higgs-sector models, the ATLAS/CMS diboson excesses are due to a neutral heavy
Higgs boson, $H'$. Large couplings of this Higgs to first-generation quarks
are assumed in order to explain the observed production rates. To accommodate
this, $H'$ must have no (appreciable) vacuum expectation value. Large
light-quark couplings to $H'$ may also lead to its production via gluon
fusion (GGF).

In strong-dynamics models, there is an approximate isospin symmetry. In some
models, there is also a left-right parity relating the masses and decay rates
of the~$\rho$ and $a$-bosons. Then, the isotriplet $\rho$-bosons' main decay
modes are {\em strong} decays into pairs of the Goldstone-boson pions which
are the longitudinal weak bosons, i.e., $\rho^\pm \to W^\pm_L Z_L$ and
$\rho^0 \to W^+_L W^-_L$. These proceed through $\rho WW$ and $\rho WZ$
interactions whose strength is nominally $g_\rho M^2_W/M^2_\rho$, where
$g_\rho$ is a strong coupling, $\simge \CO(1)$. In some models, the $\rho$
may also decay strongly into $W_L H$ and $Z_L H$. The isoscalar $\omega$ is
likely to have nearly the same mass as the $\rho$, but it prefers to decay
into $W^+_L W^-_L Z_L$, a mode not yet sought at the LHC. The same is true of
the $I=1$ axial-vectors~$a$, but there is a two-body strong decay mode
available to them, namely, $a^\pm \to W^\pm_L H$ and $a^0 \to Z_L H$. The
coupling of the $a\, W_L H$ operators is nominally $g_\rho M_W$. In
strong-dynamics models, the near degeneracy of~$\rho$ and~$a$ implied by the
parity symmetry minimizes the contribution of the strong dynamics to the
$S$-parameter~\cite{Casalbuoni:1995qt, Lane:2009ct}. The parity also forbids
$\rho \to V_L H$ and $a \to V_L V_L$ up to small electroweak
corrections~\cite{Appelquist:2015vdl}.

The coupling in the new weak gauge models is usually assumed to be of
$\CO(g)$, as for electroweak $SU(2)$. The $W'$ and $Z'$ bosons mix with their
SM counterparts, $W$ and $Z$, and this gives rise to $W'WZ$ and $Z'WW$
interactions which are nominally of order $g M^2_W/M^2_{W'}$. (A parity is at
work here too.) The only appreciable decay of the heavy $W',Z'$ to weak
bosons is to $W^\pm_L Z_L$ and $W^+_L W^-_L$. Note that neither the strong
nor this weak scenario can produce copious $ZZ$ signals.  In these models,
there often will be be $\ell^+ \ell^-$ and $\ell^\pm \nu$ signals at rates
comparable to the dibosons. The $VH$ modes occur with a large branching
fraction in the simplest LR models.

In 2HDM-inspired models, the $VH$ and $WZ$ signals come from other particles
in the heavy Higgs doublet, either $A \to VH$ or $H^{\pm} \to WZ$, though the
rates are usually larger than allowed by current data. Heavy $H'$ resonances
can also decay into $HH$, though this does not appear to have been
considered. In Refs.~\cite{Chiang:2015lqa, Cacciapaglia:2015nga}, the extended
Higgs sector consists of a neutral particle only, so no $WZ$ or $VH$ modes
are present. The scalar in this case is produced predominantly via GGF.

\subsection{Proposed Run 2 Studies}

Assuming that the ATLAS and CMS diboson excesses are confirmed in Run~2,
understanding their nature will become a major program of these
experiments. We present here a number of tests to determine whether their
origin is strong or weak dynamics, their number and electric charges, and
perhaps their spins and parities.

\begin{itemize}

\item[1)] It is very desirable that the separation of $p_T \sim 1\,\tev$,
  nonleptonically-decaying $W$ and $Z$ bosons be sharpened, and the overlap
  between nonleptonic $WW$, $WZ$ and $ZZ$ selections be minimized. In $W',Z'$
  models and strong models with isovector resonances ($\rho$, $a$), there is
  no resonant $ZZ$ production. In scalar or tensor models there are $ZZ$ and
  $WW$ resonances but, generally, very little $WZ$ signal.

\item[2)] Determine the masses and individual cross sections for each
  nonleptonic diboson mode: $VV = WW,\,WZ,\,ZZ$.

\item[3)] Discover and measure the rates of the semileptonic modes of the
  $VV$ excesses. These modes must be there if these are truly weak-boson
  resonances. Determine the ratios of semileptonic to nonleptonic rates as a
  check. In DY and VBF production of strong or weak vector bosons,
  $\sigma(\rho^+,W^{'\,+}) \simeq 3 \sigma(\rho^-,W^{'\,-})$ and
  $\sigma(\rho^\pm,W^{'\,\pm}) \simeq 2 \sigma(\rho^0,Z')$. With enough data,
  searches for 2-TeV resonant structure in all-leptonic modes will be
  possible.

\item[4)] Verify or exclude the presence of $VH = WH,\,ZH$ modes in their
  semileptonic modes (e.g., $\ell^+\nu \bar bb$ and $\ell^+ \ell^- \bar bb$,
  or leptons with $H \to \tau^+\tau^-$). Nonleptonic modes such as
  $J\bar bb$, where $J$ is a fat $V$-jet should be sought as well.

\item[5)] If $VH$ resonances exist, do they have the same mass as the $VV$
  resonances of the same electric charge? Determine individual $VH$ cross
  sections and resonance branching ratios to $VH$.

\item[6)] Measure the resonance widths in each mode. These are an important
  discriminant between strong and weak-coupling dynamics as the source of the
  resonances. The $W',\,Z'$ widths in weakly-coupled models scale as
  $g^2 M_{W',Z'}$ while the $\rho,\,a$ widths in strongly-coupled models
  scale as $g_{\rho}^2 M_{\rho,a} \gg g^2 M_{W'}$.
  
\item[7)] Discriminate between longitudinally ($V_L$) and transversely
  ($V_T$) polarized weak bosons and determine their relative fractions in the
  $VV$ and $VH$ resonances. It has been suggested that nonleptonic $V_L$ and
  $V_T$ can be distinguished by the relative $p_T$ of their decay
  subjets~\cite{Cui:2010km}. Will this be possible when
  $p_T(V) \sim 1\,\tev$? Also see Ref.~\cite{Khachatryan:2014vla}.

\item[8)] The angular distribution of the fat jets and/or lepton pairs
  relative to the beam axis will help determine the spin of the $VV$ and $VH$
  resonances. E.g., in Drell-Yan production of a heavy spin-one resonance,
  the resonance is almost at rest in the lab frame and it is polarized along
  the beam axis. Decays into $V_L V_L$ and $V_L H$ will have, approximately,
  a~$\sin^2\theta$ distribution.

\item[9)] It will be useful to discriminate between production mechanisms: DY
  vs.~VBF vs.~GGF. In VBF there will be forward jets with moderate-$p_T$ and
  a rapidity gap. Also GGF may be distinguished by the ratio of charged to
  neutral resonances. As just noted, DY production of spin-one should have
  distinctive angular distributions. The relative fraction of DY and VBF in a
  resonance's production may help determine its identity; see, e.g.,
  Ref.~\cite{Lane:2015fza}. The presence of extra jets may also indicate the
  production of multiple states or more complex decays, such as
  $W'' \to W' W \to 3W$ or $\omega \to W^+ W^- Z$; see,
  e.g.,~\cite{Aguilar-Saavedra:2015rna}.

\item[10)] Some models may have appreciable decay rates of the new resonances
  to dijets. E.g., the walking technicolor model~\cite{Fukano:2015hga}
  involves one family of technifermions and, therefore, color-octet $\rho_T$
  as well as singlets. The color-octet $\rho_T \to \bar qq,\,gg$ modes may
  have large branching ratios because, in walking technicolor, their
  $\pi_T$-pair channels are closed kinematically. In extended gauge sector
  models and multi-Higgs models such as
  Ref.~\cite{Dobrescu:2015qna,Brehmer:2015cia,Chen:2015xql,
    Omura:2015nwa} the 2-TeV resonance also has a significant branching
  fraction to dijets (either light flavor only, or both light and heavy
  flavor jets, i.e.  $t\bar
  b$).  
  
  
\item[11)] Many models predict additional bosonic resonances with masses in
  the few TeV range. Their masses and other properties are
  model-dependent. In some cases, it is possible that ``partners'' of the
  $W'$ appear before the $W'$ itself. An example is LR $W'$ models, which
  predict a~$Z'$ slightly heavier than $W'$ but which has a sizable branching
  fraction to $\ell^+ \ell^-$, facilitating its earlier detection.
  
\item[12)] The presence or absence of top (and $W,Z$) partners can also
  distinguish between certain strong-interaction models and between strongly
  and weakly-coupled models. E.g., models in which~$H$ is a pseudo-Goldstone
  boson (see, e.g., Ref.~\cite{Bellazzini:2014yua,Panico:2015jxa} for
  reviews) use such partners to stabilize the light Higgs mass, and they
  should show up soon at the LHC.

\end{itemize}

\subsection{Precision Measurements}

As explained in Section 3.4, models of the excesses involving extra Higgs
bosons often predict sizable deviations from SM of the 125~GeV Higgs'
couplings that may be observable at the LHC. 
Such deviations can be expected in some composite models with a spin-1 resonance decaying into 
electroweak bosons, particularly, but not always,  models with moderately large values of the composite 
gauge self-coupling or with largely composite light quarks.
(An exception to this is Ref.~\cite{Lane:2014vca} in which the Higgs
couplings deviate from SM by $\CO(M^2_W/M^2_\rho$).) The Higgs couplings are
more SM-like in spin-zero and spin-two models. Although a
detailed summary is beyond the scope of this report, some of those
models give large contributions to FCNC's and/or violation of lepton flavor
universality.

\section{Overview of the Models}
\label{sec:overv}

For the purpose of a condensed phenomenological overview of models addressing the diboson excess,
in Table~\ref{tab:overview} we summarize the main production modes and decay channels
of the particles responsible for the excesses in the different setups. We classify the models according to the nature of the relevant resonance(s), here spin and charges, and not according to the UV theory (which might be weakly or strongly coupled). Analyses not fitting in this simple format are listed under 'Unconventional'.

\begin{table}[h!]
\resizebox{\textwidth}{!}{
\begin{tabular}{|c||c|c|c||c|c|c|c|c|c|c|c|c|c|c|c|c||c|}
\hline
\multicolumn{18}{|c|}{{\bf Spin-1 triplets (\boldmath $V^\pm$, $V^0$)}} \\
\hline
\hline
Prod. &\!$WW$\!&$ZZ$&$WZ$&$Wh$&$Zh$&$\gamma h$&$W \gamma$&$Z \gamma$&$\gamma \gamma$&$gg$&$hh$&$\overline{Q}_3 Q_3$&$\overline{q} q$ & $l l$ & $\ell^\pm \nu$ & $X$ & Ref. \\
\hline
DY & \checkmark &   & \checkmark & & & & & & & & & (\checkmark) & (\checkmark) & (\checkmark) & (\checkmark) & & \cite{Fukano:2015hga,Cacciapaglia:2015eea,Terazawa:2015bsa,Dobado:2015hha}  \\
\hline
DY & \checkmark &   & \checkmark & \checkmark & \checkmark & & & & & & & \checkmark$_{\!\!\text{\tiny$\bar qq$}}$ & \checkmark & (\checkmark) & (\checkmark) & & \cite{Franzosi:2015zra,Abe:2015jra,Carmona:2015xaa,Bian:2015ota} \\
\hline
DY & \checkmark &   & \checkmark & \checkmark & \checkmark & & & & & & & (\checkmark) & (\checkmark) & (\checkmark) & (\checkmark) & & \cite{Lane:2015fza} \\
\hline
DY & \checkmark &   & \checkmark & \checkmark & \checkmark & & & & & & & \checkmark$_{\!\!\text{\tiny$\bar qq$}}$ & \checkmark & (\checkmark) & (\checkmark) &  (\checkmark) & \cite{Abe:2015uaa} \\
\hline
DY & \checkmark &   & \checkmark & \checkmark$_{\!\!\text{\tiny$WZ$}}$& \checkmark$_{\!\!\text{\tiny$WW$}}$ & & & & & & & \checkmark$_{\!\!\text{\tiny$\bar qq$}}$ & \checkmark & (\checkmark) & (\checkmark) & & \cite{Cheung:2015nha,Allanach:2015hba,Low:2015uha,Niehoff:2015iaa} \\
\hline
DY & \checkmark &   & \checkmark & \checkmark$_{\!\!\text{\tiny$WZ$}}$ & \checkmark$_{\!\!\text{\tiny$WW$}}$ & & & & & & & \checkmark & \checkmark & (\checkmark) & (\checkmark) & & \cite{Thamm:2015csa} \\
\hline
\hline
\multicolumn{18}{|c|}{{\bf Spin-1 \boldmath$V^0$}} \\
\hline
\hline
Prod. &\!$WW$\!&$ZZ$&$WZ$&$Wh$&$Zh$&$\gamma h$&$W \gamma$&$Z \gamma$&$\gamma \gamma$&$gg$&$hh$&$\overline{Q}_3 Q_3$&$\overline{q} q$ &  $l l$ & $\ell^\pm \nu$ & $X$ & Ref. \\
\hline
DY & \checkmark &   & & & \checkmark$_{\!\!\text{\tiny$WW$}}$ & & & & & & &  \checkmark$_{\!\!\text{\tiny$\bar qq$}}$ & \checkmark & & & & \cite{Hisano:2015gna} \\
\hline
DY & \checkmark &   & & & \checkmark$_{\!\!\text{\tiny$WW$}}$ & & & & & & &  \checkmark$_{\!\!\text{\tiny$\bar qq$}}$ & \checkmark &\checkmark  & & & \cite{Alves:2015mua}  \\
\hline
DY &  \checkmark  & \checkmark  &   &  & &  & & \checkmark & & &  &   & \checkmark & & & & \cite{Anchordoqui:2015uea}\\
\hline
\hline
\multicolumn{18}{|c|}{{\bf Spin-1 \boldmath$V^\pm$}} \\
\hline
\hline
Prod. &\!$WW$\!&$ZZ$&$WZ$&$Wh$&$Zh$&$\gamma h$&$W \gamma$&$Z \gamma$&$\gamma \gamma$&$gg$&$hh$&$\overline{Q}_3 Q_3$&$\overline{q} q$ &  $l l$ & $\ell^\pm \nu$ & $X$ & Ref. \\
\hline
DY & &   & \checkmark & \checkmark$_{\!\!\text{\tiny$WZ$}}$& & & & & & & & \checkmark$_{\!\!\text{\tiny$\bar qq$}}$ & \checkmark &  & & \checkmark & \cite{Dobrescu:2015qna,Heeck:2015qra,Dobrescu:2015yba,Dev:2015pga,Coloma:2015una} \\
\hline
DY & &   & \checkmark & \checkmark$_{\!\!\text{\tiny$WZ$}}$& & & & & & & & \checkmark$_{\!\!\text{\tiny$\bar qq$}}$ & \checkmark &  & & & \cite{Gao:2015irw,Brehmer:2015cia}\\
\hline
\hline
\multicolumn{18}{|c|}{{\bf Scalar}} \\
\hline
\hline
Prod. &\!$WW$\!&$ZZ$&$WZ$&$Wh$&$Zh$&$\gamma h$&$W \gamma$&$Z \gamma$&$\gamma \gamma$&$gg$&$hh$&$\overline{Q}_3 Q_3$&$\overline{q} q$ &  $l l$ & $\ell^\pm \nu$ & $X$ & Ref. \\
\hline
gg & \checkmark  & \checkmark   &   &  & &  & & \checkmark & \checkmark & \checkmark &  &   &  & & & & \cite{Chiang:2015lqa,Cacciapaglia:2015nga,Petersson:2015rza} \\
\hline
gg & \checkmark  & \checkmark   &   &  & &  & & (\checkmark) & (\checkmark) & \checkmark & \checkmark$_{\!\!\text{\tiny $WW/2$}}$ & (\checkmark)  &  & & & & \cite{Sanz:2015zha} \\
\hline
gg & \checkmark & \checkmark$_{\!\!\text{\tiny $WW/2$}}$  &   &  &  & \checkmark  &  &  & \checkmark & \checkmark & \checkmark &  \checkmark &  & & & (\checkmark) & \cite{Terazawa:2015bsa} \\
\hline
$q \bar{q}$ & \checkmark & \checkmark$_{\!\!\text{\tiny $WW/2$}}$  &   & (\checkmark)  & (\checkmark) &  &  &  &  &  & \checkmark &  & \checkmark  & & & \checkmark & \cite{Chen:2015xql,Omura:2015nwa,Chao:2015eea} \\
\hline
\hline
\multicolumn{18}{|c|}{{\bf 'Unconventional'}} \\
\hline
\hline
\multicolumn{17}{|c|}{Torsion-free Einstein-Cartan theory} & \cite{Xue:2015wha}   \\
\hline
\multicolumn{17}{|c|}{Tri-boson interpretation: $pp\to R \to VY \to V V^\prime X$} & \cite{Aguilar-Saavedra:2015rna} \\
\hline
\multicolumn{17}{|c|}{[Implications in other observables (direct and indirect)]} & \cite{Fukano:2015uga,Liew:2015osa,Arnan:2015csa,Jezo:2015rha, Dobado:2015hha,Deppisch:2015cua,Awasthi:2015ota} \\
\hline
\multicolumn{17}{|c|}{[Next to leading order predictions]} & \cite{Jezo:2015rha} \\
\hline
\multicolumn{17}{|c|}{[Analysis techniques]} & \cite{Goncalves:2015yua,Aguilar-Saavedra:2015yza,Llanes-Estrada:2015hfa,Bian:2015hda} \\
\hline
\end{tabular}
}
\caption{Overview of the models. 
Checkmarks highlight relevant decay channels in the model at hand,
while parentheses denote channels of subleading phenomenological importance. 
A subscript on the checkmark \checkmark$_{\!\!\text{\tiny$f$}}$ signals the branching ratio of the channel with final state $f$ to be equal (to leading order) to the one considered in that column.
We note that in some scenarios of the \textbf{Scalar} section, spin-2 resonance(s) could also be relevant for the excess (see e.g.~\cite{Sanz:2015zha}).
\label{tab:overview}}
\end{table}

\newpage

\section{Conclusions}

In this paper we have summarized the experimental situation of the alleged diboson excess around $\sim$ 2 TeV seen by the LHC. We have provided a thorough analysis of the different channels with their relative significances. We have given different theoretical interpretations that necessarily imply the existence of new resonance(s) and an extension in the sector that breaks electroweak symmetry. All models fall into either supposing a strong coupling origin for the electroweak symmetry breaking or extending the gauge or Higgs sectors. We have given an overview of the different proposals and discussed the production cross section of the resonance capable of explaining the excess and its different decay modes. In the coming days the first results will be coming from the LHC run-II and we will get further hints to whether this is a real excess or just a statistical fluctuation.

\section*{Acknowledgements}
We would like to heartily thank the organizers of the 2015 Les Houches workshop
``Physics at TeV Colliders'' where this work was initiated.  
J.\,B.\ is supported by the German Research Foundation (DFG) as part of GRK 1940 and FOR 2239.
G.B.~is supported by the National Science Foundation (grant NSF PHY-1404209).
A.C.~has been supported by a Marie Sk\l{}odowska-Curie Individual Fellowship of the European Commission's Horizon 2020 Programme under contract number 659239 NP4theLHC14.
R.S.C., K.M., and E.H.S.~are supported in part by the National Science Foundation under Grant Nos. PHY-0854889 and 1519045 to Michigan State University, as well as Grant No. PHY-1066293 to the Aspen
Center for Physics.
A.D.'s research is funded by the National Science Foundation (grant NSF PHY-1520966).
F.G.~acknowledges the support of a Marie Curie Intra European Fellowship within the EU FP7 (grant no. PIEF-GA-2013-628224). 
The work of J.L.~Hewett and T.G.~Rizzo was supported by the Department of Energy, Contract DE-AC02-76SF00515. 
The work of J.K.~ is supported by the German Research Foundation (DFG)
under grant numbers FOR~2239 and KO~4820/1--1 and by the European
Research Council (ERC) under the European Union's Horizon 2020 research
and innovation programme (grant agreement No.\ 637506).
K.L.'s research is supported in part by the U.S.~Department of
Energy under Grant No.~DE-SC0010106. 
K.L.~also gratefully acknowledges support of his Les Houches activities by the CERN Theory Group and by the Labex ENIGMASS of CNRS. 
The work of A.M. was partially supported by the National Science Foundation under Grant No. PHY-1417118.
J.M.N.~is supported by the People Programme (Marie curie Actions) of the European Union Seventh 
Framework Programme (FP7/2007-2013) under REA grant agreement PIEF-GA-2013-625809.
A.O.~is supported by grant MIURFIRB RBFR12H1MW.
C.P.~is supported by the UK Science and Technology Facilities Council (STFC) under grant ST/K001205/1.
The work of M.Q.~is partly supported by the Spanish
Consolider-Ingenio 2010 Programme CPAN (CSD2007-00042), by MINECO
under Grants CICYT-FEDER-FPA2011-25948 and
CICYT-FEDER-FPA2014-55613-P, by the Severo Ochoa Excellence Program of
MINECO under the grant SO-2012-0234, by Secretaria d'Universitats i
Recerca del Departament d'Economia i Coneixement de la Generalitat de
Catalunya under Grant 2014 SGR 1450,  and by CNPq PVE fellowship project 405559/2013-5, Brazil.
J.S. is supported by MINECO, under grant
numbers FPA2010-17915 and FPA2013-47836-C3-2-P, 
by the European Commission through the contract
PITN-GA-2012-316704 (HIGGSTOOLS) and by Junta de Andaluc\'{\i}a grants
FQM 101 and FQM 6552. 
V.S.~is supported by the STFC.
J.\,T.\ acknowledges support of the German Research Foundation (DFG) under grant number FOR 2239.

\bibliography{refs}
\bibliographystyle{utcaps}

\end{document}